\documentclass[prd,onecolumn,showpacs,floatfix,amsmath,amssymb,floatfix]{revtex4}
\usepackage{graphicx,dcolumn,booktabs,bm}
\usepackage{longtable,lscape}
\usepackage{txfonts}
\usepackage{overpic}
\usepackage{multirow}
\usepackage{amssymb}
\usepackage{indentfirst}
\usepackage{feynmf}   %{feynmp}
\usepackage{slashed}  %for Feynman symbols
\usepackage{cases}
\usepackage{color,ulem}
\usepackage{graphicx}
\usepackage[colorlinks,
            citecolor=blue,
            anchorcolor=red,
            menucolor=red,
            linkcolor=red,
            filecolor=red,
            runcolor=red,
            urlcolor=blue,
            frenchlinks=red]{hyperref}

\graphicspath{{Figures/}} % ÉèÖÃͼÐÎÎļþµÄËÑË÷·¾¶

\begin{document}
%\begin{CJK}{GBK}{

\title{Magnetic moments of the hidden-charm pentaquark states}
\author{Guang-Juan Wang$^{1}$}\email{wgj@pku.edu.cn}
\author{Rui Chen$^{2,3}$}\email{chenr2012@lzu.edu.cn}
\author{Li Ma$^{1}$}\email{lima@pku.edu.cn}
\author{Xiang Liu$^{2,3}$}\email{xiangliu@lzu.edu.cn}
\author{Shi-Lin Zhu$^{1,4}$}\email{zhusl@pku.edu.cn}

\affiliation{
$^1$Department of Physics and State Key Laboratory of Nuclear Physics and Technology and Center of High Energy Physics, Peking University, Beijing 100871, China\\
$^2$Research Center for Hadron and CSR Physics, Lanzhou University and Institute of Modern Physics of CAS, Lanzhou 730000, China\\
$^3$School of Physical Science and Technology, Lanzhou University, Lanzhou 730000, China\\
$^4$Collaborative Innovation Center of Quantum Matter, Beijing
100871, China }
%$^5$Center of High Energy Physics, Peking University, Beijing
%100871, China}

\date{\today}                                     % Activate to display a given date or no date

\begin{abstract}

The magnetic moment of a baryon state is an equally important
dynamical observable as its mass, which encodes crucial information
of its underlying structure. According to the different color-flavor
structure, we have calculated the magnetic moments of the
hidden-charm pentaquark states with the isospin
$(I,I_3)=(\frac{1}{2},\frac{1}{2})$ and $J^P={\frac{1}{2}}^{\pm}$,
${\frac{3}{2}}^{\pm}$, ${\frac{5}{2}}^{\pm}$, ${\frac{7}{2}}^{+}$ in
the molecular model, the diquark-triquark model, and the
diquark-diquark-antiquark model, respectively. Although a good
description for the pentaquark mass spectrum and decay patterns has
been obtained in all three models, different color-flavor
structures lead to different magnetic moments, which can be used to
pin down their inner structures and distinguish various models.

\end{abstract}


\maketitle

%%%%%%%%%%%%%%%%%%%%%%%%%%%%
\section{Introduction}\label{sec1}
%%%%%%%%%%%%%%%%%%%%%%%%%%%%

The LHCb Collaboration discovered two candidates of the hidden-charm
pentaquark states in the mass spectrum of $J/\psi p$
\cite{Aaij:2015tga}. According to their decay channel, there should
be five quarks $\bar c cuud$ within these two $P^{+}_c$ states. The
mass and width of the heavier structure are $(4449.8\pm1.7\pm2.5)$
MeV and $(39\pm5\pm19)$ MeV, respectively. The lighter signal has a
mass $(4380\pm8\pm29)$ MeV and a width $(205\pm8\pm6)$ MeV
\cite{Aaij:2015tga}. These two structures have opposite parity.
Their $J^P$ quantum numbers may be
$(\frac{3}{2}^{+},{\frac{5}{2}^{-}})$,
$(\frac{3}{2}^{-},{\frac{5}{2}^{+}})$ and
$(\frac{5}{2}^{-},{\frac{3}{2}^{+}})$ \cite{Aaij:2015tga}. Several
different $J^P$ combinations of the two states all give a good
description of the experiment data. Very recently, LHCb confirms
again confirmed two observed $P_c$ states with a model-independent
analysis \cite{Aaij:2016phn}. Before the observation of two $P_c$
states, the predictions of hidden-charm pentaquark were given in
Refs. \cite{Yang:2011wz,Wu:2010jy,Wu:2010vk,Wang:2011rga}.

This observation has inspired theorists to explore the $P_c$ states
extensively. Many different theoretical schemes were proposed to
interpret their inner structures. An extensive review of the current
experimental and theoretical progress on hidden-charm pentaquarks
can be found in Ref. \cite{Chen:2016qju}. These studies can be
categorized into four groups:
\begin{enumerate}
\item Molecular state scheme: since the masses of the two $P^+_c$ states are close to the mass
threshold of $\bar D^{(\ast)} \Sigma_c^{(\ast)}$, they are
speculated to be the loosely bound molecular states of $\bar
D^{(\ast)} \Sigma_c^{(\ast)}$
\cite{Chen:2015loa,Chen:2015moa,Karliner:2015ina,Huang:2015uda,Roca:2015dva,
He:2015cea,Yang:2015bmv,
Burns:2015dwa,Lu:2016nnt,Chen:2016heh,Tazimi:2016hsv,Feijoo:2015kts,Kahana:2015tkb,Shen:2016tzq}.
In fact, the existence of the molecular states composed of a heavy
baryon and a heavy meson was predicted within the one boson exchange
model before the LHCb's discovery in Ref. \cite{Yang:2011wz}. The
hidden-charm pentaquark states were also predicted and dynamically
generated with the coupled channel unitary approach in Ref.
\cite{Wu:2010jy}.

Within the framework of the molecular scheme, the mass spectrum of
the pentaquark states have been calculated with QCD sum rule
\cite{Chen:2015moa}, the quark delocalization model
\cite{Huang:2015uda}, and meson exchange model
\cite{Chen:2015loa,He:2015cea}, respectively. The strong decay ratios
of the pentaquark states have been calculated using heavy quark
symmetry \cite{Wang:2015qlf}. In the molecular model, the mass
difference between the two $P_c$ states may come from
$(m_{\Sigma_c^{\ast}}-m_{\Sigma_c})$ and the small decay width for
the heavy pentaquark states may be due to the suppression of the
P-wave decay.

\item Diquark-diquark-antiqurk scheme: Besides the molecular scheme, the $P_c$ states could have other
color-flavor configurations. For example, they were proposed as
pentaquarks the diquark-diquark-antiquark $\bar c(c q)(qq)$ structure
\cite{Maiani:2015vwa, Anisovich:2015cia, Wang:2015epa, Li:2015gta}.
In Ref. \cite{Maiani:2015vwa}, the authors suggested that the
different total spin of the light diquark and the different orbital
excitation leads to give the $70$ MeV mass difference of the two
$P_c$ states. The small decay width of the heavier state is also due
to the P-wave decay suppression. The mass spectrum was also
calculated under this ansatz \cite{Anisovich:2015cia, Wang:2015epa}.
In Ref. \cite{Li:2015gta}, the authors also calculated the week
decay ratios.

\item Diquark-triquark scheme: In Ref. \cite{Lebed:2015tna}, Lebed discussed the diquark-triquark
configuration of the pentaquark state and implications for their
decay behavior. The mass spectrum of the $P_c$ states were
calculated using the diquark-triqaurk model in \cite{Zhu:2015bba},
The possible partner states of these two $P_c$ states and their
interesting decay modes were also discussed in Refs.
\cite{Chen:2015sxa,Lebed:2015dca,Lu:2015fva}.

\item Additionally, these two structures, especially the higher one, were also suggested to
arise from the threshold effects
\cite{Liu:2015fea,Mikhasenko:2015vca,Guo:2015umn}. The $P_c$ states
were discussed in the context of the soliton picture
\cite{Scoccola:2015nia}.

\end{enumerate}

We notice that a good description for the pentaquark mass spectrum
and decay patterns has been obtained in all three models. In
other words, the mass and decay width alone can not distinguish the
inner structure of the hidden-charm pentaquarks. Recall that the
magnetic moment is an equally important dynamical observable of a
baryon state. As shown in the following sections, the different
color-flavor configurations of the $P_c$ states will lead to their
different magnetic moments, which can be used to pin down their
underlying structures.

It's the textbook knowledge that the nonrelativistic quark model is
very successful in the description of the magnetic moments of the
octet baryon ground states. We employ this simple framework to
calculate the magnetic moments of the hidden-charm pentaquark states
with different $J^P$ and color-flavor configurations.

In Sec. \ref{sec2}, we first discuss the color structure of the
different color-flavor configurations. In Sec. \ref{sec3}, we
calculate the magnetic moment of the $P_c$ states with three flavor
representations in the molecular model. In Sec. \ref{sec4}, we
calculate the $P_c$ magnetic moment in the diquark-diquark-antiquark
model. In Sec. \ref{sec5}, we derive the $P_c$ magnetic moment
with the diquark-triquark model. In Sec. \ref{sec6}, we discuss
the numerical results and summarize.

%%%%%%%%%%%%%%%%%%%%%%%%%%%%%%%%%
\section{Color configuration of the $P_c$ states}
\label{sec2}
%%%%%%%%%%%%%%%%%%

According to the color configurations, the pentaquark state may be
composed of either two or three clusters. For the two-cluster case,
one cluster contains two quarks while the other cluster contains
three quarks. According to the color structures of the two clusters,
the pentaquark state may be a molecular state or a diquark-triquark
state. Their possible configurations are $(\bar c c)(q_1q_2q_3)$,
$(\bar c q_3)(c q_1q_2)$, $(c q_3)(\bar c q_1q_2)$ and
$(q_1q_2)(\bar c c q_3)$. The pentaquark state may also contain
three clusters: two diquarks plus an antiquark. The configuration is
$(cq_3)(q_1q_2)\bar c$, $(\bar c q_3)(q_1 q_2)c$ and $(c \bar
c)(q_1q_2) q_3$.

For illustration, we first define the so-called color factor $I$ of
the cluster which indicates whether there exists attraction within
the cluster.
\begin{equation}
I= \sum_{i<j, a=1}^{8} \lambda_i^a \lambda_j^a ={1\over 2}\left(
C[p,q]-n C[1,0]\right),
\end{equation}
where $\lambda_i^a$ is the color generator of the ith quark with the
cluster, $n=2, 3$ for the diquark and triquark, respectively, and
$C[p,q]={p^2+pq+q^2\over 3}+p+q$ is the first-rank Casimir operator
for the cluster with the color respresentation $[p,q]$. For example,
$C[1,0]=C[0,1]={4\over 3}$ for the quark and antiquark.

For either the $(\bar c q_3)$ or ($\bar c c$)cluster within the
configuration $(\bar c c)(q_1q_2q_3)$ and $(\bar c q_3)(q_1q_2)$, we
have $\bar 3\otimes 3=1\oplus8$. In other words, the color factor of
the $(\bar c q_3)$ or ($\bar c c$)cluster is $-\frac{4}{3}$ for the
color singlet and $\frac{1}{6}$ for the color octet.

The color representation for the three particle cluster is
$3\otimes(3\otimes 3)=3\otimes(\bar 3\oplus
6)=(1\oplus8_2)\oplus(10\oplus8_1)$ with the color factor
$I_{color}=(-2, -{1\over 2}, 1, -{1\over 2})$. To form a color
singlet $P_c$ state, the diquark and triquark should combine as
follows:
\begin{eqnarray} \label{ww1}
 1\otimes1&=&1,\nonumber\\
 8\otimes8_1&=&1\oplus 8\oplus 8\oplus 10\oplus \overline{10}\oplus 27 ,\nonumber\\
8\otimes8_2&=&1\oplus 8\oplus 8\oplus 10\oplus \overline{10}\oplus
27 .\nonumber
\end{eqnarray}

In order to form a quasibound cluster, either the $(\bar c q_3)$ or
($\bar c c$) tends to be the color singlet. In other words, the
$(\bar c q_3)$ forms a heavy meson and ($\bar c c$) becomes a
charmonium. Accordingly, the $c q_1q_2$ or $q_1q_2 q_3$ cluster
forms a color singlet baryon. Since there does not exist strong
attraction between a charmonium and a light baryon, it's hard for
them to form a loosely bound molecular state. We do not consider
this possibility.

In contrast, the $(\bar c q_3)(c q_1q_2)$ configuration could lead
to the hidden-charm molecular states composed of $\bar
D^{(\ast)}\Sigma_c^{(\ast)}$, which have been studied extensively
\cite{Yang:2011wz,Wu:2010jy,Chen:2015loa,Roca:2015dva}. Therefore,
we will consider the $(\bar c q_3)(c q_1q_2)$ structure.

For the diquark within the configurations $(c q_3)(\bar c q_1q_2)$
and $(q_1q_2)(\bar c c q_3)$, we have $3\otimes 3=\bar 3\oplus 6$
with the color factor $I=-\frac{2}{3}, \frac{1}{3}$, respectively.
For the triquark, $\bar 3\otimes (3\otimes3)=\bar 3\otimes (\bar
3\oplus6)=(\bar 3\otimes \bar 3 )\oplus (\bar3 \otimes6)=(3_1\oplus
\bar 6)\oplus(3_2\oplus 15)$ with the color factor $
I=\frac{1}{3}(-4,-1,-4,2)$.

The triquark with the $3_1$ and $3_2$ representation tends to form a
quasibound cluster with strong attraction, which then couples with
the diquark in the $\bar 3$ representation to form a color singlet
pentaquark state.

Now let us move on to the diquark-diquark-antiquark model. We note
that this model converts into the molecular state if there exists
one cluster with a $\bar c$ and a quark. The color representation of
this cluster prefers the color singlet after the decomposition
$3\otimes \bar 3=1\oplus 8$. Then the other three quarks form
another color singlet. The discussion is the same as in the case of
a molecular state. In other words, the only nontrivial
configuration in the diquark-diquark-antiquark model is
$(cq_3)(q_1q_2)\bar c$, where both of the two diquarks are in the
$\bar 3_c$ representation.

In the above, we have used the color algebra to discuss the
molecular configuration $(\bar c q_3)(c q_1q_2)$, the
diquark-diquark-antiquark configuration $\bar c (cq_3)(q_1q_2)$ and
diquark-triquark configurations $(c q_3)(\bar c q_1q_2)$ and
$(q_1q_2)(\bar c c q_3)$. In the following sections, we will use
these configurations to construct the color, flavor, spin and
orbital wave functions of the $P_c$ states, which will be used in
the calculation of their magnetic moments. As will be shown in the
subsequent sections, different models with different color
representations lead to different magnetic moments.

%%%%%%%%%%%%%%%%%%%%%%%%%%%%%%%%%%%%%%%
\section{Magnetic moment of the $P_c$ states in the molecular scheme}
\label{sec3}
%%%%%%%%%%%%%%%%%%%%%%

As a molecular state, $\bar c q_3$ and $cq_1q_2$ are in the color
singlet. The two hadrons form a loosely bound molecular state.
$q_1q_2$ inside the baryon should be in ${\bar 3}_c$ representation
according to Sec. \ref{sec2}. We study the molecular state in a
$SU(3)_f\times SU(3)_c$ frame. Due to the Fermi statistics, there
are two different configurations for $q_1q_2$ as listed in Table
\ref{c1}. As the heavy quark does not carry isospin number, we have
only considered the three light quarks in constructing the flavor
wave functions.

\renewcommand{\arraystretch}{1.8}
\begin{table}[htbp]
\caption{Structure of $q_1q_2$ in the baryon under the molecular
frame. Here, S and A represent the wave function of the
corresponding space are symmetry and antisymmetry, respectively.
}\label{c1}
\begin{tabular}{c|ccc|c|cccc}
 \toprule[1pt]
\multirow{4}{*}{$J_{baryon}=\frac{1}{2}$}& color & ${\bar3}_c$& A &\multirow{4}{*}{$J_{baryon}=\frac{1}{2},\frac{3}{2}$}& color & ${\bar3}_c$& A \\
%\hline
 & space  & $L=0$ &  S&& space  & $L=0$ &  S  \\
%\hline
 &spin  & $s=0$ &  A & &spin  & $s=1$ &  S \\
%\hline
 &flavor  & $\bar{3}_f$ &  A &&flavor  & $6_f$ &  S \\
\bottomrule[1pt]
\end{tabular}
\end{table}

\begin{figure}[htbp]
  \centering
  \includegraphics[width=0.3\hsize]{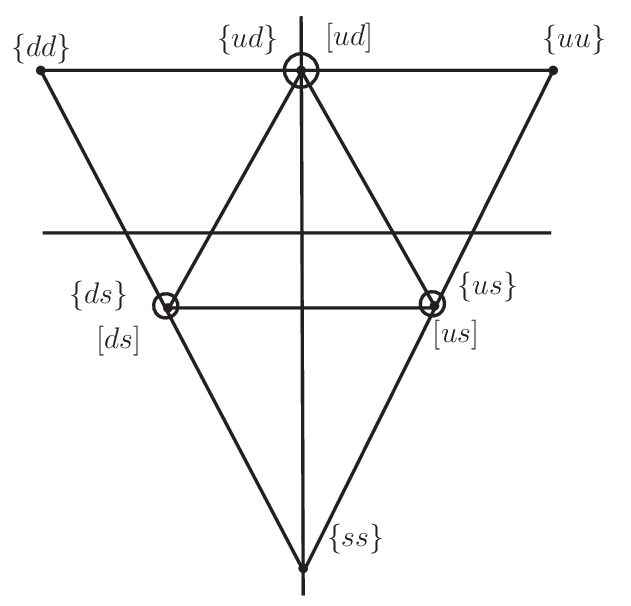}\\
  \caption{The $q_1q_2$ forms the $6_f  {\oplus}\bar {3}_{f}$ flavor representation.
  The $\{q_1q_2\}=\frac{1}{\sqrt{2}}(q_1q_2+q_2q_1)$ is for the symmetric $6_f$
  representation, while the $[q_1q_2] =\frac{1}{\sqrt{2}}(q_1q_2-q_2q_1)$ belongs
  to the antisymmetric $ \bar 3_f$ representation. The three circles represent
  the $\bar 3_f$ representation.}\label{f1}
  \end{figure}

\begin{figure}[htbp]
  \centering
  \includegraphics[width=0.5\hsize]{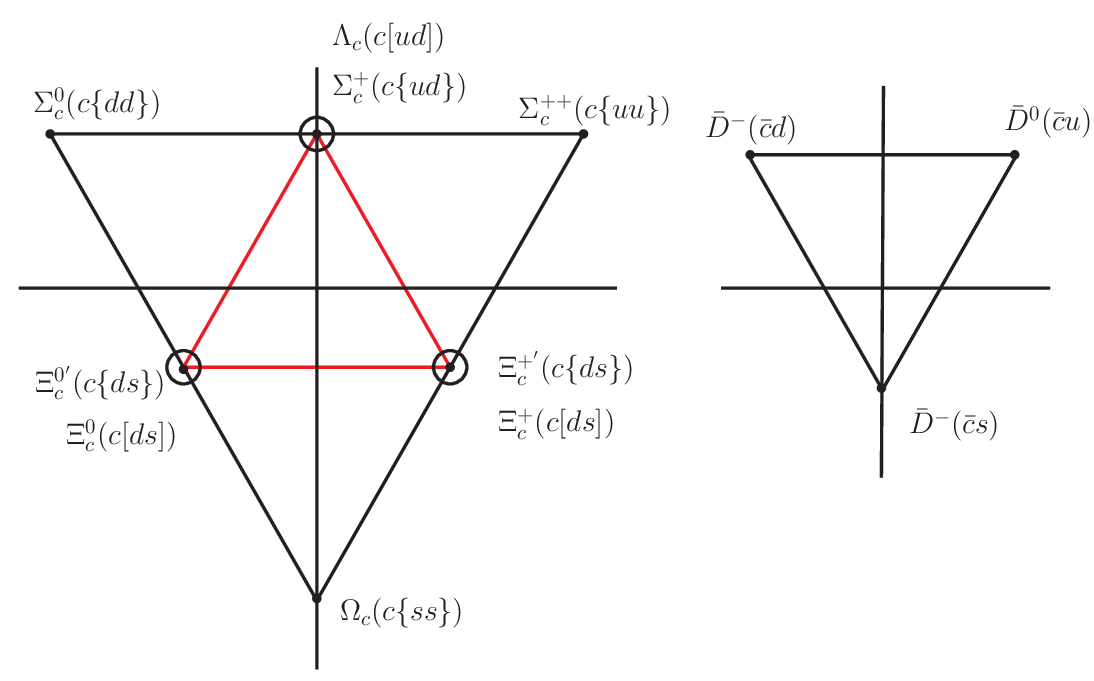}\\
  \caption{The $\bar 3_f\oplus 6_f $ and $3_f$ representations of the two clusters
  $(c q_1q_2)$ and $(\bar c q_3)$, respectively. Here, $[q_1q_2]=\frac{1}{\sqrt 2}(q_1q_2-q_2q_1)$,
  $\{q_1q_2\}=\frac{1}{ \sqrt 2}(q_1q_2+q_2q_1)$. }\label{f2}
  \end{figure}

\begin{figure}[htbp]
  \centering
  \includegraphics[width=0.5\hsize]{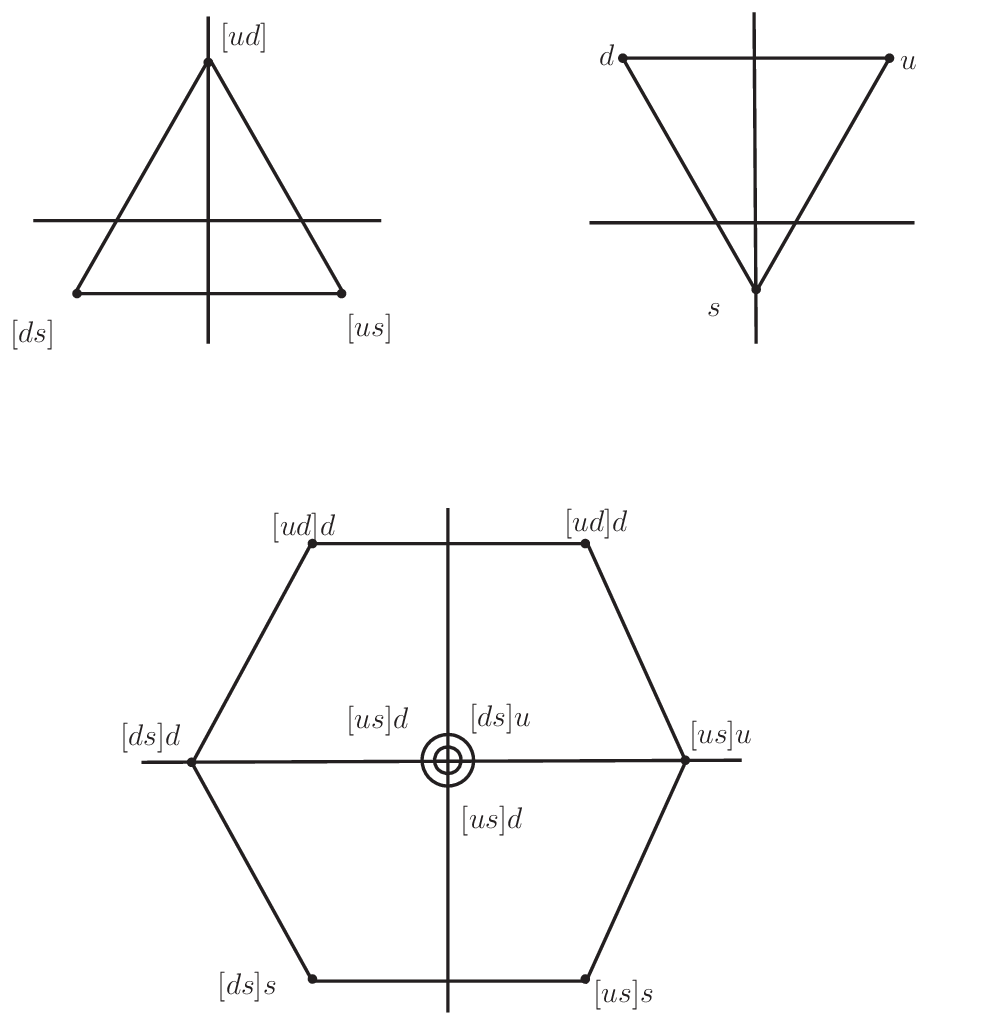}\\
  \caption{The $[q_1q_2]$ in the $\bar3_f$ representation combines with the $q_3$ to form
  the flavor combination ${1_f}{\oplus}8_{2f}$. Here, $[q_1q_2] =\frac{1}{\sqrt{2}}(q_1q_2-q_2q_1)$.} \label{f3}
     \end{figure}

\begin{figure}[htbp]
  \centering
  \includegraphics[width=0.3\hsize]{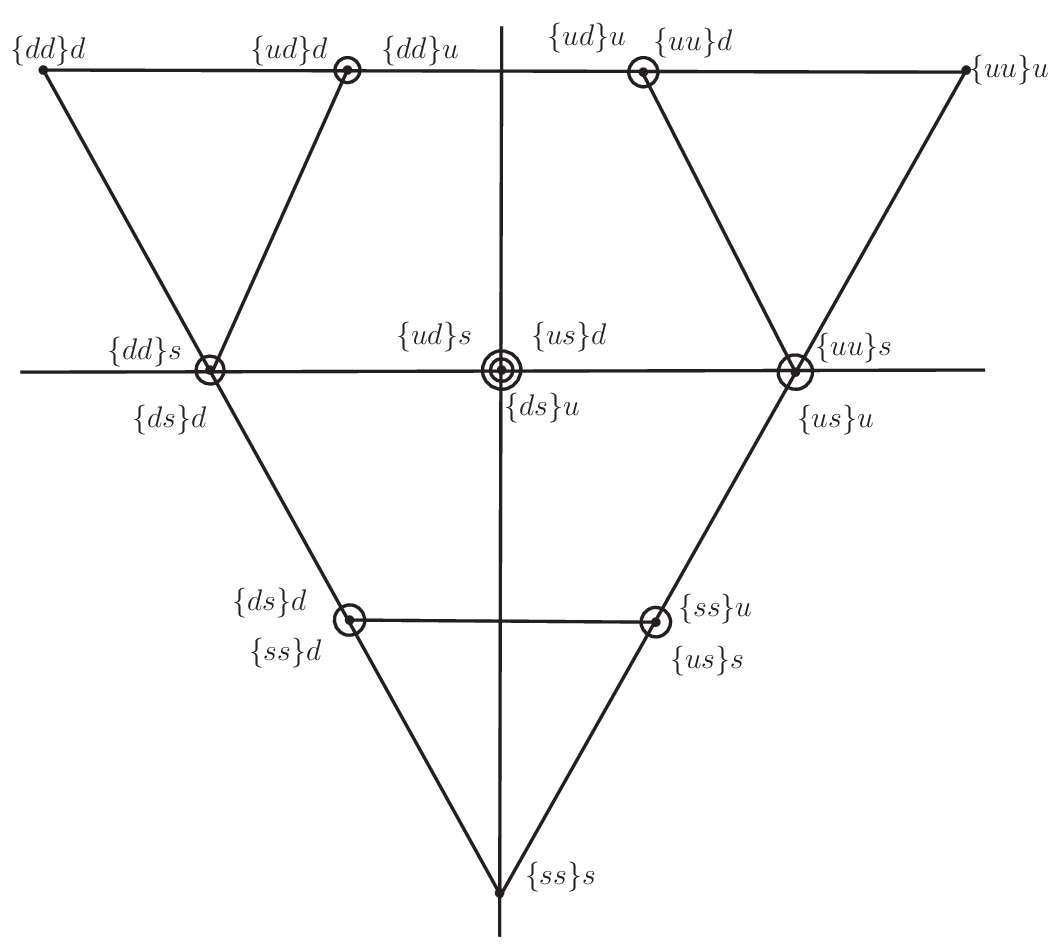}\\
  \caption{The $\{q_1q_2\} $ in the $6_f$ representation combines with the $q_3$ to form the
  ${8_{1f}}{\oplus}10_f$. Here, $\{q_1q_2\} =\frac{1}{\sqrt{2}}(q_1q_2+q_2q_1)$. Here, the seven circles and
  one circle in the origin indicate the $8_{1f}$ representation.}\label{f4}
     \end{figure}

The $q_1q_2$ forms the $\bar 3_f$ and $ 6_f$ flavor representation
with the total spin $S=0$ and $1$, respectively. The two flavor
representations are listed in Fig. \ref{f1}. In Fig. \ref{f2}, we
illustrate the heavy baryons in the $\bar 3_f$ and $6_f$
representation. In Table \ref{c1}, $[q_1q_2]$ forms the ${\bar3}_f$,
which then combines with the $q_3$ to form the flavor representation
$\bar{3}_f\otimes 3_f={1_f} \oplus 8_{2f}$ as illustrated in Fig.
\ref{f3}. When the $q_1q_2$ is in the $6_f$,
$6_f\otimes{3_f}=10_f\otimes{8_{1f}}$ as listed in Fig. \ref{f4}.
After taking into account the Clebsch-Gordan coefficients, we get
the flavor wave function of the pentaquark state with the
configuration $(\bar c q_3)(cq_1q_2)$. The same method is applied to
the other configurations $(c q_3)(\bar c q_1 q_2)$ and $(c
q_3)(q_1q_2) \bar c$. The wave functions of the pentaquark states
are listed in Table \ref{10-flavor}.

\renewcommand{\arraystretch}{1.8}
\begin{table*}[htbp]
\caption{The flavor wave functions of the molecular pentaquark
states with the configuration $(\bar c q_3)(cq_1q_2)$ in different
representations from ${3}_f\otimes{3}_f\otimes
3_f=10_f\oplus8_{1f}\oplus8_{2f}\oplus1_f$. Here, $8_{1f}$ and
$8_{2f}$ representation correspond to the $6_f$ and $\bar{3}_f$
representation of $q_1q_2$, respectively.}\label{10-flavor}
\begin{center}
   \begin{tabular}{c|c|c} \toprule[1pt]\toprule[1pt]
 $(Y, I, I_3)$ &\multicolumn{1}{c|} {Wave function-$8_{1f}$}   &Wave function-$8_{2f}$\\
\midrule[1pt]

$(1,\frac{1}{2},\frac{1}{2})$ &$-\sqrt{\frac{1}{3}}({\bar
c}u)(c\{ud\})+\sqrt{\frac{2}{3}}({\bar c}d)(c\{uu\})$
& $({\bar c}u)(c[ud])$\\
$(1,\frac{1}{2},-\frac{1}{2})$  &$\sqrt{\frac{1}{3}}({\bar
c}d)(c\{ud\})-\sqrt{\frac{2}{3}}({\bar c}u)(c\{dd\})$
&  $({\bar c}d)(c[ud])$ \\
$(-1,\frac{1}{2},\frac{1}{2})$ &$\sqrt{\frac{1}{3}}({\bar
c}s)(c\{us\})-\sqrt{\frac{2}{3}}({\bar c}u)(c\{ss\})$
&  $({\bar c}s)(c[us])$  \\
$(-1,\frac{1}{2},-\frac{1}{2})$  &$\sqrt{\frac{1}{3}}({\bar
c}s)(c\{ds\})-\sqrt{\frac{2}{3}}({\bar c}d)(c\{ss\})$
&        $({\bar c}s)(c[ds])$ \\
$ (0,1,1)$& $\sqrt{\frac{1}{3}}({\bar
c}u)(c\{us\})-\sqrt{\frac{2}{3}}({\bar c}s)(c\{uu\})$
&  $({\bar c}u)(c[us])$\\
$ (0,1,0)$& $\sqrt{\frac{1}{6}}[({\bar c}d)(c\{us\})+({\bar
c}u)(c\{ds\})]-\sqrt{\frac{2}{3}}({\bar c}s)(c\{ud\})$
& $\frac{1}{\sqrt2}\{ ({\bar c}d)(c[us])+({\bar c}u)(c[ds]) \}$ \\
$ (0,1,-1)$& $\sqrt{\frac{1}{3}}({\bar
c}d)(c\{ds\})-\sqrt{\frac{2}{3}}({\bar c}s)(c\{dd\})$
& $({\bar c}d)(c[ds])$\\
$ (0,0,0)$& $\sqrt{\frac{1}{2}}[({\bar c}u)(c\{ds\})-({\bar
c}d)(c\{us\})]$
&    $\frac{1}{\sqrt6} \{ ({\bar c}d)(c[us])-({\bar c}u)(c[ds])-2(\bar c s)(c[ud]) \}$  \\
     \bottomrule[1pt]

$(Y, I, I_3)$ &\multicolumn{1}{c|} {Wave function-$10_f$}
&Singlet\\\midrule[1pt] $(1,\frac{3}{2},\frac{3}{2})$& $({\bar
c}u)(c\{uu\})$
&   $\frac{1}{\sqrt3} \{ ({\bar c}d)(c[us])-({\bar c}u)(c[ds])+(\bar c s)(c[ud]) \}$ \\
$(1,\frac{3}{2},\frac{1}{2})$  &  $\sqrt{\frac{2}{3}}({\bar c}u)(c\{ud\})+\sqrt{\frac{1}{3}}({\bar c}d)(c\{uu\})$  \\
$(1,\frac{3}{2},-\frac{1}{2})$&  $\sqrt{\frac{2}{3}}({\bar c}d)(c\{ud\})+\sqrt{\frac{1}{3}}({\bar c}u)(c\{dd\})$  \\
$(1,\frac{3}{2},-\frac{3}{2})$&  $({\bar c}d)(c\{dd\})$ \\
$ (0,1,1)$& $\sqrt{\frac{2}{3}}({\bar c}u)(c\{us\})+\sqrt{\frac{1}{3}}({\bar c}s)(c\{uu\})$       \\
$ (0,1,0)$& $\sqrt{\frac{1}{3}}[({\bar c}d)(c\{us\})+({\bar c}u)(c\{ds\})]+\sqrt{\frac{1}{3}}({\bar c}s)(c\{ud\})$        \\
$ (0,1,-1)$& $\sqrt{\frac{2}{3}}({\bar c}d)(c\{ds\})+\sqrt{\frac{1}{3}}({\bar c}s)(c\{dd\})$        \\
$ (-1,\frac{1}{2},\frac{1}{2})$& $\sqrt{\frac{1}{3}}({\bar c}u)(c\{ss\})+\sqrt{\frac{2}{3}}({\bar c}s)(c\{us\})$        \\
$ (-1,\frac{1}{2},-\frac{1}{2})$& $\sqrt{\frac{1}{3}}({\bar c}d)(c\{ss\})+\sqrt{\frac{2}{3}}({\bar c}s)(c\{ds\})$          \\
$(-2,0,0)$&  $({\bar c}s)(c\{ss\})$  \\
     \bottomrule[1pt]\bottomrule[1pt]
      \end{tabular}
  \end{center}
\end{table*}

Within the molecular scheme, there is a P-wave orbital excitation in
the state with positive parity. The P-wave excitation inside the
baryon or meson would increase the mass of the pentaquark by $400$
MeV. Moreover, a P-wave heavy hadron would be broad and unstable,
which would render the pentaquark very broad. Such a
possibility is not favored by the LHCb's observation. Therefore, the
orbital excitation lies between the heavy meson and heavy baryon for
the $P_c$ state with positive parity in the molecular scheme.

The magnetic moment of the pentaquark with the configuration $(\bar
c q_3)(cq_1q_2)$ is \cite{Liu:2003ab}
\begin{eqnarray} \label{w1}
 \vec{\mu}&=&g_{m}{\mu_{m}}{\vec{S}_m}+g_{b}{\mu_b}{\vec{S}}_b+{\mu_l}\vec{l} ,\nonumber\\
g_{m}{\mu_{m}}{\vec{S}_m}&=&g_{\bar Q} {\mu_{\bar Q}}\vec{\frac{1}{2}}+g_{q_3}{\mu_{q_3}}\vec{\frac{1}{2}}, \nonumber\\
 g_{b}{\mu_b}{\vec{S}}_b&=&g_{Q} {\mu_Q}\vec{\frac{1}{2}}+g_{q_1}{\mu_{q_1}}\vec{\frac{1}{2}}+g_{q_2}{\mu_{q_2}}\vec{\frac{1}{2}},\nonumber\\
{\mu_l}&=&\frac{{m_b}\mu_{m}+{m_m}\mu_{b}}{m_m+m_b},
\end{eqnarray}
where $g$ is the Lande factor. The subscripts $m$ and $b$ represent
meson and baryon, respectively. $l$ is the orbital excitation
between the two hadrons. $Q$ and $\bar Q$ are the heavy quark and
antiquark. $q_{1,2,3}$ is the light quark. $m_{m(b)}$ is the meson
(or baryon) mass. $\mu$ denotes the magnetic moment. The pentaquark
state is written as
\begin{eqnarray} \label{w2}
|JJ\rangle=\left|\left[\left[\left((q_1q_2)_{s_{12}}c_{\frac{1}{2}}\right)_{S_b}\otimes
(\bar cq_3)_{S_m}\right]_S\otimes l\right]_{J}^{J}\right\rangle,
\end{eqnarray}
\begin{eqnarray} \label{w3}
\mu &=&\left\langle JJ\left |g_{m}{\mu_{m}}{\vec{S}_m}+g_{b}{\mu_b}{\vec{S}}_b+{\mu_l}\vec{l}\,\,\right|JJ\right\rangle \nonumber\\
&=&\sum_{l_z, {S_z}}\left\langle l l_z,SS_z|JJ\right\rangle^2
\left \{ \mu_{l}l_z+\sum_{S'_z}\left\langle {S_b}S'_z,S_m (S_z-S'_z)|SS_z\right \rangle^2\left[ (S_z-S'_z)(\mu_{\bar Q}+\mu_{q_3})\right.\right.\nonumber\\
&&\left.\left.+\left\langle
{\frac{1}{2}}{\frac{1}{2}},s_{12}\left(S'_z-\frac{1}{2}\right)\bigg|S_b
S'_z\right\rangle^2\left(\mu_{\bar
Q}+\left(S'_z-\frac{1}{2}\right)(\mu_{q_1}+\mu_{q_2})\right)+\left\langle
{\frac{1}{2}}{-\frac{1}{2}},s_{12}(S'_z+\frac{1}{2})\bigg|S_b
S'_z\right\rangle^2\left(-\mu_{Q}+\left(S'_z+\frac{1}{2}\right)\left(\mu_{q_1}+\mu_{q_2}\right)\right)\right]
\right\},\nonumber\\
\end{eqnarray}
where $s_{12}$ and $S$ are the spin of the diquark inside the baryon
and the spin of the pentaquark state, respectively. $l_z$, $S_z$, and
$S'_z$ are the third component of the orbital excitation, the total
spin of the pentaquark state, the spin of the baryon, respectively.

The values of the constituent quark masses are
\begin{eqnarray} \label{p1}
m_s=0.5 \,\mbox{GeV},\  m_u=m_d=0.33 \,\mbox{GeV}, \ m_c=1.55
\,\mbox{GeV},
\end{eqnarray}
which lead to the magnetic moments of the octet baryons in rough
agreement with the experimental data as listed in Table
\ref{baryon}.

\renewcommand{\arraystretch}{1.8}
\begin{table}[htbp]
\caption{The magnetic moment of the octet baryons. The unit is the
magnetic moment of the proton.} \label{baryon}
\begin{center}
\begin{tabular}{c|c|c|c}
 \toprule[1pt]\toprule[1pt]
Baryons & Magnetic moment& Numerical& Experiment  \\
\midrule[1pt]
$p$ & ${\frac{4}{3}{\mu_u}-\frac{1}{3}{\mu}_d}$  & $2.842$ &  $2.793$  \\
\hline
$n$ & $\frac{4}{3}{\mu}_d-\frac{1}{3}{\mu}_u$  & $-1.895$&$-1.913$  \\
\hline
$\Lambda$ & ${\mu}_s$ & $-0.625$&$-0.613\pm0.006$ \\
\hline
$\Sigma^{+}$ & $\frac{4}{3}{{\mu}_u}-\frac{1}{3}{{\mu}_s}$  & $2.735$ & $2.460\pm0.006$   \\
\hline
$\Sigma^{-}$ & $\frac{4}{3}{\mu}_d-\frac{1}{3}{\mu}_s$  & $-1.055$ & $-1.160\pm0.025$   \\
\hline
$\Xi^{0}$ & $\frac{4}{3}{\mu}_s-\frac{1}{3}{\mu}_u$  & $-1.465$ & $-1.250\pm0.014$   \\
\hline
$\Xi^{-}$ & $\frac{4}{3}{\mu}_s-\frac{1}{3}{\mu}_d$  & $-0.518$ & $-0.651\pm0.0025$  \\
\hline
$\Omega^{-}$ & $3{\mu}_s$  & $-1.876$ & $-2.020\pm0.05$   \\
\bottomrule[1pt]\bottomrule[1pt]
\end{tabular}
\end{center}
\end{table}

We use the masses of the heavy mesons and baryons from PDG
\cite{Agashe:2014kda}. Within the molecular scheme, the $P_c(4380)$
and $P_c(4450)$ state with isospin $\frac{1}{2}$ sit right on the
mass threshold of $\bar D^{(\ast)}\Sigma_c^{(\ast)}$ and belong to
the $8_{1f}$ representation since they were observed in the $J/\psi
p$ channel. Their magnetic moments are listed in Table \ref{pcm}.
The unit is the proton magnetic moment $\mu_p$.

\renewcommand{\arraystretch}{1.5}
\begin{table*}[htbp]
\caption{The magnetic moment of the molecular pentaquark states $(\bar c q_3)(cq_1q_2)$ with %$(\bar c q_3)(cq_1q_2)$
the isospin $(I,I_3)=(\frac{1}{2},\frac{1}{2})$ in the $8_{1f}$
flavor representation from ${6}_f\otimes 3_f=10_f\oplus8_{1f}$. The third line denotes the $J^P$ of the heavy baryon, meson, and orbital excitation. The
unit is the magnetic moment of the proton.}\label{pcm}
 %\scriptsize
\begin{center}
   \begin{tabular}{c|c|c|c|c|c|c|c} \toprule[1pt]\toprule[1pt]

&\multicolumn{3}{c|}{$^{2}S_{\frac{1}{2}}$
($J^P={\frac{1}{2}}^{-}$)}
&\multicolumn{3}{c|}{${^{4}S_{\frac{3}{2}}}$
(${J^P={\frac{3}{2}}^{-}}$)}
&\multicolumn{1}{c}{$^{6}S_{\frac{5}{2}}^{-}$
($J^P={\frac{5}{2}}^{-}$)} \\\cline{2-8} $(Y, I, I_3)$ &
${\frac{1}{2}}^{+}\otimes0^{-}\otimes0^{+}$
      & ${\frac{1}{2}}^{+}\otimes1^{-}\otimes0^{+}$  &  ${\frac{3}{2}}^{+}\otimes1^{-}\otimes0^{+}$      & ${\frac{1}{2}}^{+}\otimes1^{-}\otimes0^{+}$ &${\frac{3}{2}}^{+}\otimes0^{-}\otimes0^{+}$
      & ${\frac{3}{2}}^{+}\otimes1^{-}\otimes0^{+}$  & ${\frac{3}{2}}^{+}\otimes1^{-}\otimes0^{+}$
\\
\hline

$(1,\frac{1}{2},\frac{1}{2})$ &1.760 &-0.856 &1.938&1.357 &3.246 &2.219 &2.842 \\
   \bottomrule[1pt]

 &\multicolumn{2}{c|}{${^{2}P_{\frac{1}{2}}}$ (${J^P={\frac{1}{2}}^{+}}$)} &\multicolumn{2}{c|}{${^{4}P_{\frac{1}{2}}}$ (${J^P={\frac{1}{2}}^{+}}$)} &\multicolumn{1}{c|}{${^{2}P_{\frac{1}{2}}}$ (${J^P={\frac{1}{2}}^{+}}$)} &\multicolumn{1}{c|}{${^{4}P_{\frac{1}{2}}}$ (${J^P={\frac{1}{2}}^{+}}$)}  \\\cline{2-8}
$(Y, I, I_3)$ & ${\frac{1}{2}}^{+}\otimes0^{-}\otimes1^{-}$  &  $[{\frac{1}{2}}^{+}\otimes1^{-}]_{\frac{1}{2}}\otimes1^{-}$      & $[{\frac{1}{2}}^{+}\otimes1^{-}]_{\frac{3}{2}}\otimes1^{-}$ &${\frac{3}{2}}^{+}\otimes0^{-}\otimes1^{-}$    & $[{\frac{3}{2}}^{+}\otimes1^{-}]_{\frac{1}{2}}\otimes1^{-}$   & $[{\frac{3}{2}}^{+}\otimes1^{-}]_{\frac{3}{2}}\otimes1^{-}$\\
\hline
$(1,\frac{1}{2},\frac{1}{2})$ &-0.530 &0.363 &0.715&1.779 &-0.578 &1.199  \\

 \bottomrule[1pt]

&\multicolumn{2}{c|}{${^{2}P_{\frac{3}{2}}}$
(${J^P={\frac{3}{2}}^{+}}$)}
&\multicolumn{2}{c|}{${^{4}P_{\frac{3}{2}}}$
(${J^P={\frac{3}{2}}^{+}}$)}
&\multicolumn{1}{c|}{${^{2}P_{\frac{3}{2}}}$
(${J^P={\frac{3}{2}}^{+}}$)}
&\multicolumn{1}{c|}{${^{4}P_{\frac{3}{2}}}$
(${J^P={\frac{3}{2}}^{+}}$)}
&\multicolumn{1}{c}{${^{6}P_{\frac{3}{2}}}$
(${J^P={\frac{3}{2}}^{+}}$)}  \\\cline{2-8}
$(Y, I, I_3)$ & ${\frac{1}{2}}^{+}\otimes0^{-}\otimes1^{-}$  &  $[{\frac{1}{2}}^{+}\otimes1^{-}]_{\frac{1}{2}}\otimes1^{-}$      & $[{\frac{1}{2}}^{+}\otimes1^{-}]_{\frac{3}{2}}\otimes1^{-}$ &${\frac{3}{2}}^{+}\otimes0^{-}\otimes1^{-}$    & $[{\frac{3}{2}}^{+}\otimes1^{-}]_{\frac{1}{2}}\otimes1^{-}$   & $[{\frac{3}{2}}^{+}\otimes1^{-}]_{\frac{3}{2}}\otimes1^{-}$   & $[{\frac{3}{2}}^{+}\otimes1^{-}]_{\frac{5}{2}}\otimes1^{-}$\\
\hline

$(1,\frac{1}{2},\frac{1}{2})$  &1.846 &-0.740 &1.041      &2.409 &2.040 &1.668 &2.326 \\

     \bottomrule[1pt]

 &\multicolumn{3}{c|}{${^{4}P_{\frac{5}{2}}}$ (${J^P={\frac{5}{2}}^{+}}$)} &\multicolumn{1}{c|}{${^{6}P_{\frac{5}{2}}}$ (${J^P={\frac{5}{2}}^{+}}$)}  &\multicolumn{1}{c|}{${^{6}P_{\frac{7}{2}}}$ (${J^P={\frac{7}{2}}^{+}}$)}\\\cline{2-8}
$ (Y, I, I_3)$   &${\frac{1}{2}}^{+}\otimes1^{-}\otimes1^{-}$ &${\frac{3}{2}}^{+}\otimes0^{-}\otimes1^{-}$    & $[{\frac{3}{2}}^{+}\otimes1^{-}]_{\frac{3}{2}}\otimes1^{-}$   & $[{\frac{3}{2}}^{+}\otimes1^{-}]_{\frac{5}{2}}\otimes1^{-}$   & ${\frac{3}{2}}^{+}\otimes1^{-}\otimes1^{-}$\\
\hline

$(1,\frac{1}{2},\frac{1}{2})$  &1.473 &3.319 &2.322      &2.547 &2.945 \\
    \bottomrule[1pt]\bottomrule[1pt]
      \end{tabular}
  \end{center}
\end{table*}

The molecular states $P_c(4380)$ and $P_c(4450)$ have the flavor
wave function as $-\sqrt{\frac{1}{3}}\Sigma_c^{+(*)}\bar
D^{0(*)}+\sqrt{\frac{2}{3}}\Sigma_c^{++(*)}\bar D^{-(*)}$.
% Their masses approach the mass threshold of $\Sigma_c^{(\ast)}\bar D^{(\ast)}$.
 For the molecular state composed of $\Sigma_c\bar D^{\ast}$ with $J^P=\frac{3}{2}^{+}$, there are two configurations $[\frac{1}{2}^{+}\otimes1^{-}]_{\frac{1}{2}}\otimes 1^{-}$ and $[\frac{1}{2}^{+}\otimes1^{-}]_{\frac{3}{2}}\otimes 1^{-}$. Their magnetic moments are quite different, i.e. $-0.740\mu_p$ and $1.041\mu_p$, respectively. Therefore, the study of the magnetic moment can help probe the inner structures of the $P_c$ state. For the $\Sigma_c^{\ast} \bar D^{\ast}$ state with $J^P=\frac{3}{2}^{\pm},\frac{5}{2}^{\pm}$, the magnetic moments
are around $2$. For example, the $\Sigma_c^{\ast}\bar D^{\ast}$
state with  $J^P=\frac{3}{2}^{+}$ has three different
configurations, $[\frac{3}{2}^{+}\otimes1^{-}]_{\frac{1}{2}}\otimes
1^{-}$, $[\frac{3}{2}^{+}\otimes1^{-}]_{\frac{3}{2}}\otimes 1^{-}$
and $[\frac{3}{2}^{+}\otimes1^{-}]_{\frac{5}{2}}\otimes 1^{-}$.
Their magnetic moments are $2.040 \mu_p$, $1.668 \mu_p$ and $2.326
\mu_p$, respectively.

%%%%%%%%%%%%%%%%%%%%%%%%%%%%%%
\section{Magnetic moment of the $P_c$ states in the diquark-diquark-antiquark model}
\label{sec4}
%%%%%%%%%%%%%%%%%%%%%%%%%%%%

The diquark-diquark-antiquark configuration is $(cq_3)(q_1q_2)\bar
c$. The magnetic moment of the $P_c$ state with positive parity
depends on the position of the P-wave excitation. Similar to the
P-wave light baryon, there are two P-wave excitation modes inside
the three-body bound state, the $\rho$ and the $\lambda$ excitation.
The $\rho$ mode P-wave orbital excitation lies between the diquark
$c q_3$ and diquark $q_1q_2$. The $\lambda$ mode P-wave orbital
excitation lies between the $\bar c$ and the center of mass system
of the $c q_3$ and $q_1q_2$. Now the pentaquark state $|JJ\rangle$
is written as
\begin{eqnarray} \label{w4}
|JJ\rangle=\left|\left[\left[\left[(cq_3)_{s_1}\otimes(q_1q_2)_{s_2}\right]_{s'}\otimes
{\bar{c}}_{\frac{1}{2}}\right]_S\otimes l\right]_J^{J}\right\rangle
,
\end{eqnarray}
where $s_1$ and $s_2$ are the spin of the diquark $cq_3$ and
$q_1q_2$, respectively. They two couple into the spin $s'$, which
combines with the spin of the $\bar c$ to form the total spin of the
pentaquark state $S$. $l$ is the orbital excitation. $J$ is the
total angular momentum. The magnetic moment of the pentaquark state
is
\begin{eqnarray} \label{w5}
\mu &=&\left\langle JJ|g_{1}{\mu_{1}}{\vec{S}_1}+g_{2}{\mu_2}{\vec{S}}_2+g_{\bar Q}{\mu_{\bar Q}}\vec{\frac{1}{2}}+{\mu_l}\vec{l}\,\,\bigg|JJ\right\rangle \nonumber\\
&=&\sum_{m, {S_z}}\langle lm,SS_z|JJ\rangle^2
\left \{ \mu_{l}m+\sum_{S'_z}\left\langle S'S'_z \frac{1}{2}(S_z-S'_z)\bigg|SS_z\right\rangle^2\left[2(S_z-S'_z){\mu_{\bar Q}}\right.\right. \nonumber\\
&&\left.\left.+\sum_{{S_1}_z}\left\langle
S_1{S_1}_zS_2(S'_z-{S_1}_z)\bigg|S'S'_z\right\rangle^2\left({S_1}_z(\mu_Q+\mu_{q3})+(S'_z-{S_1}_z)(\mu_{q1}+{\mu_{q2}})\right)\right]\right\},
\end{eqnarray}
where $m$, $S_z$, and $S'_z$ are the third component of the orbital
excitation, the total spin of the pentaquark state, and the diquark
spin, respectively. We use the values of the diquark masses from
Ref. \cite{Ebert:2010af}
\begin{eqnarray} \label{p2}
&&m_{[q q]}=0.710\,\mbox{GeV},  \ m_{\{qq\}}=0.909\,\mbox{GeV},\ m_{[qs]}=0.948\,\mbox{GeV},\ m_{ \{ qs \} }=1.069\,\mbox{GeV},\ m_{\{ss\}}=1.203\,\mbox{GeV},  \nonumber\\
&&m_{[c q]}=1.937\,\mbox{GeV},  \ m_{\{cq\}}=2.036\,\mbox{GeV},\
m_{[cs]}=2.091\,\mbox{GeV},\ m_{ \{ cs \} }=2.158\,\mbox{GeV}.
\end{eqnarray}

We list the flavor wave functions of the pentaquark states in this
model in Table \ref{dda8-flavor}. The analytic expressions for
magnetic moments of the configurations $(cq_3)(q_1q_2)\bar c$ in the
$8_{2f}$ and $8_{1f}$ flavor representations are in Table
\ref{analysis3}. In Tables \ref{dda8r} and \ref{dda8l}, we list the
magnetic moments of the pentaquark states with the isospin
$(I,I_3)=(\frac{1}{2},\frac{1}{2})$. The states with the positive
parities have the $\rho$ and $\lambda$ excitation modes in the two
tables, respectively.
\renewcommand{\arraystretch}{1.8}
 \begin{table*}[htbp]
 \caption{The flavor wave functions of the pentaquark states in the
 diquark-diquark-antiquark model with the configuration $(c
 q_3)(q_1q_2)\bar c$ in different representations from ${3}_f\otimes{3}_f\otimes
3_f=10_f\oplus8_{1f}\oplus8_{2f}\oplus1_f$. Here, $8_{1f}$ and
$8_{2f}$ representation correspond to the representation for
$q_1q_2$ being $6_f$ and $\bar{3}_f$,
respectively.}\label{dda8-flavor}
  \begin{center}
   \begin{tabular}{c|c|c} \toprule[1pt]\toprule[1pt]

$(Y, I, I_3)$ &\multicolumn{1}{c|} {Wave function-$8_{1f}$}
&Wave function-$8_{2f}$\\
 \midrule[1pt]
 $(1,\frac{1}{2},\frac{1}{2})$ &$-\sqrt{\frac{1}{3}}(cu)\{ud\}{\bar c}+\sqrt{\frac{2}{3}}({c}d)\{uu\}{\bar c}$
 & $({c}u)[ud]\bar c$\\
 $(1,\frac{1}{2},-\frac{1}{2})$  &$\sqrt{\frac{1}{3}}(cd)c\{ud\}{\bar c}-\sqrt{\frac{2}{3}}({c}u)\{dd\}{\bar c}$
 &  $(cd)[ud]{\bar c}$\\
$ (0,1,1)$& $\sqrt{\frac{1}{3}}({c}u)\{us\}{\bar
c}-\sqrt{\frac{2}{3}}(cs)\{uu\}{\bar c}$
&   $(cu)[us]{\bar c}$ \\

$ (0,1,0)$& $\sqrt{\frac{1}{6}}[({c}d)\{us\}{\bar
c}+({c}u)\{ds\}{\bar c}]-\sqrt{\frac{2}{3}}({c}s)\{ud\}{\bar c}$
& $\frac{1}{\sqrt2}\{ (cd)[us]{\bar c}+(cu)[ds]{\bar c} \}$ \\
$ (0,1,-1)$& $\sqrt{\frac{1}{3}}({c}d)\{ds\}{\bar
c}-\sqrt{\frac{2}{3}}({c}s)\{dd\}{\bar c}$
& $(cd)[ds]{\bar c}$  \\
$ (0,0,0)$& $\sqrt{\frac{1}{2}}[({c}u)\{ds\}{\bar
c}-({c}d)\{us\}{\bar c}]$
&    $\frac{1}{\sqrt6} \{ (cd)[us]{\bar c}-(cu)[ds]{\bar c}-2(cs)[ud]{\bar c} \}$\\
$(-1,\frac{1}{2},\frac{1}{2})$ &$\sqrt{\frac{1}{3}}({c}s)\{us\}{\bar
c}-\sqrt{\frac{2}{3}}({c}u)\{ss\}{\bar c}$
&  $(cs)[us]{\bar c}$ \\
$(-1,\frac{1}{2},-\frac{1}{2})$
&$\sqrt{\frac{1}{3}}({c}s)\{ds\}{\bar
c}-\sqrt{\frac{2}{3}}({c}d)\{ss\}{\bar c}$
&        $(cs)[ds]{\bar c}$ \\
\bottomrule[1pt]

$(Y, I, I_3)$ &\multicolumn{1}{c|} {Wave function-$10_f$}
&Singlet\\
\midrule[1pt]

$(1,\frac{3}{2},\frac{3}{2})$& $({c}u)\{uu\}{\bar c}$
&   $\frac{1}{\sqrt3} \{ ({c}d[us]){\bar c}-(cu)[ds]{\bar c}+(cs)[ud]{\bar c} \}$ \\

$(1,\frac{3}{2},\frac{1}{2})$  &  $\sqrt{\frac{2}{3}}({c}u)\{ud\}{\bar c}+\sqrt{\frac{1}{3}}({c}d)\{uu\}{\bar c}$ \\
$(1,\frac{3}{2},-\frac{1}{2})$&  $\sqrt{\frac{2}{3}}({c}d)\{ud\}{\bar c}+\sqrt{\frac{1}{3}}({c}u)\{dd\}{\bar c}$  \\
$(1,\frac{3}{2},-\frac{3}{2})$&  $({c}d)(\{dd\}){\bar c}$ \\
$ (0,1,1)$& $\sqrt{\frac{2}{3}}({c}u)\{us\}{\bar c}+\sqrt{\frac{1}{3}}({c}s)\{uu\}{\bar c}$\\
$ (0,1,0)$& $\sqrt{\frac{1}{3}}[({c}d)\{us\}{\bar c}+({c}u)\{ds\}{\bar c}]+\sqrt{\frac{1}{3}}({c}s)\{ud\}{\bar c}$\\
$ (0,1,-1)$& $\sqrt{\frac{2}{3}}({c}d)\{ds\}{\bar c}+\sqrt{\frac{1}{3}}({c}s)\{dd\}{\bar c}$\\
$ (-1,\frac{1}{2},\frac{1}{2})$& $\sqrt{\frac{1}{3}}({c}u)\{ss\}{\bar c}+\sqrt{\frac{2}{3}}({c}s)\{us\}{\bar c}$ \\
$ (-1,\frac{1}{2},-\frac{1}{2})$& $\sqrt{\frac{1}{3}}({c}d)\{ss\}{\bar c}+\sqrt{\frac{2}{3}}({c}s)\{ds\}{\bar c}$  \\
$(-2,0,0)$&  $({c}s)\{ss\}{\bar c}$   \\
     \bottomrule[1pt]\bottomrule[1pt]
     \end{tabular}
 \end{center}
\end{table*}

\renewcommand{\arraystretch}{1.6}
\begin{table*}[htbp]
%\scriptsize
\caption{The analytic formula for the magnetic moment of the
pentaquark in the configuration $(Q q_3)(q_1q_2)\bar Q$ when
$s_{q_1q_2}=0$ and $1$.}\label{analysis3}
\begin{center}
   \begin{tabular}{c|c|c|c} \toprule[1pt]\toprule[1pt]
 \multicolumn{4}{c} {$s_{q_1q_2}=0$}\\\toprule[1pt]
$J^P$ &$^{2s+1}L_J$ & &Formula \\\midrule[1pt]

{$\frac{1}{2}^{-}$}  &{{${^{2}S_{\frac{1}{2}}}$}}   &${0}^{+}\otimes0^{+}\otimes{\frac{1}{2}}^{-} \otimes{0^{+}}$ &$\mu _{\bar{Q}}$\\

& &${1}^{+}\otimes0^{+}\otimes{\frac{1}{2}}^{-} \otimes{0^{+}}$ &$\frac{1}{3} \left(2 \mu _Q+2 \mu _{{q_3}}-\mu _{\bar{Q}}\right)$\\

\hline
{$\frac{3}{2}^{-}$}  &{{${^{4}S_{\frac{3}{2}}}$}}   & ${1}^{+}\otimes0^{+}\otimes{\frac{1}{2}}^{-} \otimes{0^{+}}$  &$\mu _Q+\mu _{{q_3}}+\mu _{\bar{Q}}$\\
\hline

{$\frac{1}{2}^{+}$}  &{{${^{2}P_{\frac{1}{2}}}$}}  & ${0}^{+}\otimes0^{+}\otimes{\frac{1}{2}}^{-} \otimes{1^{-}}$ & $\frac{1}{3} \left(2 \mu_1-\mu_{\bar{Q}}\right)$\\

& &$({1}^{+}\otimes0^{+}\otimes{\frac{1}{2}}^{-})_{\frac{1}{2}}
\otimes{1^{-}}$  &$\frac{1}{9} \left(6 \mu_1-2 \mu_Q-2
\mu_{{q_3}}+\mu_{\bar{Q}}\right)$\\\cline{2-4}

  &{\multirow{1}*{${^{4}P_{\frac{1}{2}}}$}}  &$({1}^{+}\otimes0^{+}\otimes{\frac{1}{2}}^{-})_{\frac{3}{2}} \otimes{1^{-}}$  &$\frac{1}{9} \left(-3 \mu _1+5 \left(\mu _Q+\mu _{{q_3}}+\mu _{\bar{Q}}\right)\right)$\\

\hline

{$\frac{3}{2}^{+}$}  &{{${^{2}P_{\frac{3}{2}}}$}}   &${0}^{+}\otimes0^{+}\otimes{\frac{1}{2}}^{-} \otimes{1^{-}}$   & $\mu _1+\mu _{\bar{Q}}$\\

&
&$({1}^{+}\otimes0^{+}\otimes{\frac{1}{2}^-})_{\frac{1}{2}}\otimes{1^{-}}$
&$\frac{1}{3} \left(3 \mu _1+2 \mu _Q+2 \mu _{{q_3}}-\mu
_{\bar{Q}}\right)$\\\cline{2-4}

  &{\multirow{1}*{${^{4}P_{\frac{3}{2}}}$}} &$({1}^{+}\otimes0^{+}\otimes{\frac{1}{2}^-} )_{\frac{3}{2}}\otimes{1^{-}}$  &$\frac{1}{15} \left(6 \mu _1+11 \left(\mu _Q+\mu _{{q_3}}+\mu _{\bar{Q}}\right)\right)$\\

\hline

\multirow{1}{*}{$\frac{5}{2}^{+}$}  &{\multirow{1}*{${^{4}P_{\frac{5}{2}}}$}}  &${1}^{+}\otimes0^{+}\otimes{\frac{1}{2}^-}\otimes{1^{-}}$   &$\mu _1+\mu _Q+\mu _{{q_3}}+\mu _{\bar{Q}}$\\
\bottomrule[1pt]
  %    \end{tabular}
 % \end{center}
%\end{table*}

%\renewcommand{\arraystretch}{1.8}
%\begin{table}[htbp]
%\caption{The analytic formula for the magnetic moment of the
%pentaquark state in the configuration $(Q q_3)(q_1q_2)\bar Q$.} \label{analysis 4-}
%  \begin{center}
%   \begin{tabular}{c|c|c|c} \toprule[1pt]\toprule[1pt]
   \multicolumn{4}{c} {$s_{q_1q_2}=1$}\\\toprule[1pt]
\,\,\,\,$J^P$\,\,\,\,    &\,$^{2s+1}L_J$\,  &   &Formula
\\\midrule[1pt]

{$\frac{1}{2}^{-}$}  &{{${^{2}S_{\frac{1}{2}}}$}}   &$0^{+}\otimes1^{+} \otimes {\frac{1}{2}}^{-}\otimes0^{+}$ &$\frac{1}{3}\left(2 \mu _{{q_1}}+2 \mu _{{q_2}}-\mu _{\bar{Q}}\right)$\\

&&$(1^{+}\otimes1^{+})_{0} \otimes {\frac{1}{2}}^{-}\otimes0^{+}$ &$\mu _{\bar{Q}}$\\
&&$(1^{+}\otimes1^{+})_{1} \otimes {\frac{1}{2}}^{-}\otimes0^{+}$  &$\frac{1}{3} \left(\mu _Q+\mu _{{q_1}}+\mu _{{q_2}}+\mu _{{q_3}}-\mu _{\bar{Q}}\right)$\\

\hline

{$\frac{3}{2}^{-}$}  &{{${^{4}S_{\frac{3}{2}}}$}}    &$(0^{+}\otimes1^{+})\otimes{\frac{1}{2}}^{-}\otimes0^{+}$  &$\mu _{{q_1}}+\mu _{{q_2}}+\mu _{\bar{Q}}$\\

&&$(1^{+}\otimes1^{+})_{1} \otimes {\frac{1}{2}}^{-}\otimes0^{+}$ &$\frac{1}{2} \left(\mu _Q+\mu _{{q_1}}+\mu _{{q_2}}+\mu _{{q_3}}+2 \mu _{\bar{Q}}\right)$\\

& &$(1^{+}\otimes1^{+})_{2} \otimes {\frac{1}{2}}^{-}\otimes0^{+}$  &$\frac{3}{10} \left(3 \mu _Q+3 \mu _{{q_1}}+3 \mu _{{q_2}}+3 \mu _{{q_3}}-2 \mu _{\bar{Q}}\right)$\\
\hline

\multirow{1}{*}{$\frac{5}{2}^{-}$}  &{\multirow{1}*{${^{6}S_{\frac{5}{2}}}$}} &$1^{+}\otimes1^{+}\otimes{\frac{1}{2}}^{-}\otimes0^{+}$  &$\mu _Q+\mu _{{q_1}}+\mu _{{q_2}}+\mu _{{q_3}}+\mu _{\bar{Q}}$\\

\hline

{$\frac{1}{2}^{+}$}  &{{${^{2}P_{\frac{1}{2}}}$}}   &$(0^{+}\otimes1^{+}\otimes {\frac{1}{2}}^{-})_{\frac{1}{2}}\otimes1^{-}$  &$\frac{1}{9} \left(6 \mu _1-2 \mu _{{q_1}}-2 \mu _{{q_2}}+\mu _{\bar{Q}}\right)$\\

&&$((1^{+}\otimes1^{+})_{0}\otimes{\frac{1}{2}}^{-})_{\frac{1}{2}}\otimes1^{-}$
&$\frac{1}{3} \left(2 \mu _1-\mu _{\bar{Q}}\right)$\\

&&$((1^{+}\otimes1^{+})_{1}\otimes{\frac{1}{2}}^{-})_{\frac{1}{2}}\otimes1^{-}$ &$\frac{1}{9} \left(6 \mu _1-\mu _Q-\mu _{{q_1}}-\mu _{{q_2}}-\mu _{{q_3}}+\mu _{\bar{Q}}\right)$\\

\cline{2-4}
 &{{${^{4}P_{\frac{1}{2}}}$}}  &$(0^{+}\otimes1^{+}\otimes {\frac{1}{2}}^{-})_{\frac{3}{2}}\otimes1^{-}$ &$\frac{1}{9} \left(-3 \mu _1+5 \left(\mu _{{q_1}}+\mu _{{q_2}}+\mu _{\bar{Q}}\right)\right)$\\

&&$((1^{+}\otimes1^{+})_{1}\otimes{\frac{1}{2}}^{-})_{\frac{3}{2}}\otimes1^{-}$   &$\frac{1}{18} \left(-6 \mu _1+5 \left(\mu _Q+\mu _{{q_1}}+\mu _{{q_2}}+\mu _{{q_3}}+2 \mu _{\bar{Q}}\right)\right)$ \\ &&$((1^{+}\otimes1^{+})_{2}\otimes{\frac{1}{2}}^{-})_{\frac{3}{2}}\otimes1^{-}$ &$\frac{1}{6} \left(-2 \mu _1+3 \mu _Q+3 \mu _{{q_1}}+3 \mu _{{q_2}}+3 \mu _{{q_3}}-2 \mu _{\bar{Q}}\right)$\\

\hline

{$\frac{3}{2}^{+}$}  &{{${^{2}P_{\frac{3}{2}}}$}}   &$(0^{+}\otimes1^{+}\otimes {\frac{1}{2}}^{-})_{\frac{1}{2}}\otimes1^{-}$  &$\frac{1}{3} \left(3 \mu _1+2 \mu _{{q_1}}+2 \mu _{{q_2}}-\mu _{\bar{Q}}\right)$\\

&&$((1^{+}\otimes1^{+})_{0}\otimes{\frac{1}{2}}^{-})_{\frac{1}{2}}\otimes1^{-}$ & $\mu _1+\mu _{\bar{Q}}$\\
&&$((1^{+}\otimes1^{+})_{1}\otimes{\frac{1}{2}}^{-})_{\frac{1}{2}}\otimes1^{-}$ & $\frac{1}{3} \left(3 \mu _1+\mu _Q+\mu _{{q_1}}+\mu _{\text{q2}}+\mu _{{q_3}}-\mu _{\bar{Q}}\right)$\\

\cline{2-4}

 &{{${^{4}P_{\frac{3}{2}}}$}}  &$(0^{+}\otimes1^{+}\otimes {\frac{1}{2}}^{-})_{\frac{3}{2}}\otimes1^{-}$ &$\frac{1}{15} \left(6 \mu _1+11 \left(\mu _{{q_1}}+\mu _{{q_2}}+\mu _{\bar{Q}}\right)\right)$\\

   &&$((1^{+}\otimes1^{+})_{1}\otimes{\frac{1}{2}}^{-})_{\frac{3}{2}}\otimes1^{-}$     &$\frac{1}{30} \left(12 \mu _1+11 \left(\mu _Q+\mu _{{q_1}}+\mu _{q_2}+\mu _{{q_3}}+2 \mu _{\bar{Q}}\right)\right)$\\
   & &$((1^{+}\otimes1^{+})_{2}\otimes{\frac{1}{2}}^{-})_{\frac{3}{2}}\otimes1^{-}$   &$\frac{1}{50} \left(20 \mu _1+11 \left(3 \mu _Q+3 \mu _{{q_1}}+3 \mu _{\text{q2}}+3 \mu _{{q_3}}-2 \mu _{\bar{Q}}\right)\right)$ \\

\cline{2-4}

&{\multirow{1}*{${^{6}P_{\frac{3}{2}}}$}} &$((1^{+}\otimes1^{+})_{2}\otimes{\frac{1}{2}}^{-})_{\frac{5}{2}}\otimes1^{-}$ &$\frac{3}{25} \left(-5 \mu _1+7 \left(\mu _Q+\mu _{{q_1}}+\mu _{{q_2}}+\mu _{{q_3}}+\mu _{\bar{Q}}\right)\right)$\\

\hline

{$\frac{5}{2}^{+}$}  &{{${^{4}P_{\frac{5}{2}}}$}}  &$(0^{+}\otimes1^{+}\otimes {\frac{1}{2}}^{-})\otimes1^{-}$ &$\mu _1+\mu _{{q_1}}+\mu _{{q_2}}+\mu _{\bar{Q}}$\\

 &&$((1^{+}\otimes1^{+})_{1}\otimes{\frac{1}{2}}^{-})_{\frac{3}{2}}\otimes1^{-}$ &$\frac{1}{2} \left(2 \mu _1+\mu _Q+\mu _{\text{q1}}+\mu _{{q_2}}+\mu _{{q_3}}+2 \mu _{\bar{Q}}\right)$\\

        &&$((1^{+}\otimes1^{+})_{2}\otimes{\frac{1}{2}}^{-})_{\frac{3}{2}}\otimes1^{-}$ &$\frac{1}{10} \left(10 \mu _1+9 \mu _Q+9 \mu _{{q_1}}+9 \mu _{\text{q2}}+9 \mu _{{q_3}}-6 \mu _{\bar{Q}}\right)$\\
\cline{2-4}

&{\multirow{1}*{${^{6}P_{\frac{5}{2}}}$}} &$((1^{+}\otimes1^{+})_{2}\otimes{\frac{1}{2}}^{-})_{\frac{5}{2}}\otimes1^{-}$ &$\frac{1}{35} \left(10 \mu _1+31 \left(\mu _Q+\mu _{{q_1}}+\mu _{\text{q2}}+\mu _{{q_3}}+\mu _{\bar{Q}}\right)\right)$\\

\hline

\multirow{1}{*}{$\frac{7}{2}^{+}$}&{\multirow{1}*{${^{6}P_{\frac{5}{2}}}$}} &$1^{+}\otimes1^{+}\otimes{\frac{1}{2}}^{-}\otimes1^{-}$ &$\mu _1+\mu _Q+\mu _{{q_1}}+\mu _{\text{q2}}+\mu _{{q_3}}+\mu _{\bar{Q}}$\\
     \bottomrule[1pt]\bottomrule[1pt]
      \end{tabular}
  \end{center}
\end{table*}

\renewcommand{\arraystretch}{1.7}
\begin{table*}[htbp]
\scriptsize \caption{The magnetic moment of the pentaquark states in
the diquark-diquark-antiquark model with the configuration $(c
q_3)(q_1q_2)\bar c$ and the isospin
$(I,I_3)=(\frac{1}{2},\frac{1}{2})$. They are in the $8_{2f}$
representation from ${\bar3}_f\otimes 3_f=1_f\oplus8_{2f}$ and
$8_{1f}$ representation from ${6}_f\otimes 3_f=10_f\oplus8_{1f}$, respectively. The third line denotes the $J^P$ of the diquark
$(cq_3)$, diquark $(q_1q_2)$, antiquark, and orbital excitation,
respectively. For the pentaquark states with positive parity, the
P-wave excitation is the $\rho$ mode. The unit is the magnetic
moment of the proton.}\label{dda8r} \tiny
\begin{center}
   \begin{tabular}{c|c|c|c|c|c|c|c} \toprule[1pt]\toprule[1pt]
 \multicolumn{8}{c} {Flavor representation-$8_{2f}$}\\\toprule[1pt]
&\multicolumn{2}{c|}{$^2S_{\frac{1}{2}}$ ($J^P={\frac{1}{2}}^{-}$)}
&\multicolumn{1}{c|}{$^{4}S{{\frac{3}{2}}}$
(${J^P={\frac{3}{2}}^{-}}$)}
&\multicolumn{2}{c|}{${^{2}P_{\frac{1}{2}}}$
(${J^P={\frac{1}{2}}^{+}}$)}
&\multicolumn{1}{c|}{${^{4}P_{\frac{1}{2}}}$
(${J^P={\frac{1}{2}}^{+}}$)} &\\\cline{2-7} $(Y, I, I_3)$ &
${0}^{+}\otimes0^{+}\otimes{\frac{1}{2}}^{-} \otimes{0^{+}}$
      & ${1}^{+}\otimes0^{+}\otimes{\frac{1}{2}}^{-} \otimes{0^{+}}$  & ${1}^{+}\otimes0^{+}\otimes{\frac{1}{2}}^{-} \otimes{0^{+}}$  &  $0^{+}\otimes0^{+}\otimes{\frac{1}{2}}^{-} \otimes{1^{-}}$ &$({1}^{+}\otimes0^{+}\otimes{\frac{1}{2}}^{-})_{\frac{1}{2}} \otimes{1^{-}}$ &$({1}^{+}\otimes0^{+}\otimes{\frac{1}{2}}^{-})_{\frac{3}{2}}\otimes{1^{-}}$ \\
\hline

$(1,\frac{1}{2},\frac{1}{2})$&-0.403&1.667&1.895&0.462&-0.232&0.913 \\

\bottomrule[1pt]

&\multicolumn{2}{c|}{$^{2}P_{\frac{3}{2}}$
($J^P={\frac{3}{2}}^{+}$)}
&\multicolumn{1}{c|}{$^{4}P_{\frac{3}{2}}$
($J^P={\frac{3}{2}}^{+}$)}
&\multicolumn{1}{c|}{${^{4}P_{\frac{5}{2}}^{+}}$
(${J^P={\frac{5}{2}}^{+}}$)}   \\\cline{2-7} \hline

$(Y, I, I_3)$ & ${0}^{+}\otimes0^{+}\otimes{\frac{1}{2}}^{-}
\otimes{1^{-}}$
      & $({1}^{+}\otimes0^{+}\otimes{\frac{1}{2}}^{-} )_{\frac{1}{2}}\otimes{1^{-}}$ & $({1}^{+}\otimes0^{+}\otimes{\frac{1}{2}}^{-} )_{\frac{3}{2}}\otimes{1^{-}}$ & ${1}^{+}\otimes0^{+}\otimes{\frac{1}{2}}^{-} \otimes{1^{-}}$ \\
\hline

$(1,\frac{1}{2},\frac{1}{2})$ &0.088&2.152&1.638&2.380  \\

\bottomrule[1pt]
%\end{tabular}
  %\end{center}
%\end{table*}
%\renewcommand{\arraystretch}{1.8}
%\begin{table*}[htbp]
%\caption{The magnetic moment of the pentaquark states in thediquark-diquark-antiquark model with the configuration $(cq_3)(q_1q_2)\bar c$ in the $8_{1f}$ flavor representation from${6}_f\otimes 3_f=10_f\oplus8_{1f}$. The third line denotes the$J^P$ of the diquark $(cq_3)$, diquark $(q_1q_2)$, anti-quark andorbital excitation respectively. For the pentaquark states withpositive parity, the $\rho$ mode P-wave excitation is between $cq_3$ and $q_1q_2$. The unit is the magnetic moment of the proton.}\label{adda8-} \tiny
%  \begin{center}
%  \begin{tabular}{c|c|c|c|c|c|cc} \toprule[1pt]\toprule[1pt]

\multicolumn{8}{c} {Flavor representation-$8_{1f}$}\\\toprule[1pt]
&\multicolumn{3}{c|}{$^{2}S_{\frac{1}{2}}$
($J^P={\frac{1}{2}}^{-}$)}
&\multicolumn{3}{c|}{${^{4}S_{\frac{3}{2}}}$
(${J^P={\frac{3}{2}}^{-}}$)} &\multicolumn{1}{c}{$^{6}
S_{\frac{5}{2}}$ ($J^P={\frac{5}{2}}^{-}$)}
\\\cline{2-8}

$(Y, I, I_3)$ & $0^{+}\otimes1^{+} \otimes
{\frac{1}{2}}^{-}\otimes0^{+}$
      &$(1^{+}\otimes1^{+})_{0} \otimes {\frac{1}{2}}^{-}\otimes0^{+}$  &$(1^{+}\otimes1^{+})_{1} \otimes {\frac{1}{2}}^{-}\otimes0^{+}$     &$(0^{+}\otimes1^{+})\otimes{\frac{1}{2}}^{-}\otimes0^{+}$ &$(1^{+}\otimes1^{+})_{1} \otimes {\frac{1}{2}}^{-}\otimes0^{+}$
       &$(1^{+}\otimes1^{+})_{2} \otimes {\frac{1}{2}}^{-}\otimes0^{+}$   &$(1^{+}\otimes1^{+})\otimes {\frac{1}{2}}^{-}\otimes0^{+}$
\\
\hline

$(1,\frac{1}{2},\frac{1}{2})$  &2.029 &-0.403 &1.216
      &2.439 &1.219 &3.163 &2.842 \\

    \bottomrule[1pt]

 &\multicolumn{1}{c|}{$^{2}P_{\frac{1}{2}}$ (${J^P={\frac{1}{2}}^{+}}$)}   &\multicolumn{1}{c|}{$^{4}P_{\frac{1}{2}}$ (${J^P={\frac{1}{2}}^{+}}$)} &\multicolumn{2}{c|}{$^{2}P_{\frac{1}{2}}$ (${J^P={\frac{1}{2}}^{+}}$)} &\multicolumn{2}{c|}{$^{4}P_{\frac{1}{2}}$ (${J^P={\frac{1}{2}}^{+}}$)}\\\cline{2-8}

$(Y, I, I_3)$         &$(0^{+}\otimes1^{+}\otimes
{\frac{1}{2}}^{-})_{\frac{1}{2}}\otimes1^{-}$
&$(0^{+}\otimes1^{+}\otimes
{\frac{1}{2}}^{-})_{\frac{3}{2}}\otimes1^{-}$
&$((1^{+}\otimes1^{+})_{0}\otimes{\frac{1}{2}}^{-})_{\frac{1}{2}}\otimes1^{-}$
         &$((1^{+}\otimes1^{+})_{1}\otimes{\frac{1}{2}}^{-})_{\frac{1}{2}}\otimes1^{-}$

               &$((1^{+}\otimes1^{+})_{1}\otimes{\frac{1}{2}}^{-})_{\frac{3}{2}}\otimes1^{-}$     &$((1^{+}\otimes1^{+})_{2}\otimes{\frac{1}{2}}^{-})_{\frac{3}{2}}\otimes1^{-}$\\
\hline

$(1,\frac{1}{2},\frac{1}{2})$ &-0.139 &1.086 &0.673
      &0.133 &0.408 &1.488  \\

     \bottomrule[1pt]

&\multicolumn{1}{c|}{${^2 P_{\frac{3}{2}}}$
(${J^P={\frac{3}{2}}^{+}}$)} &\multicolumn{1}{c|}{${^4
P_{\frac{3}{2}}}$ (${J^P={\frac{3}{2}}^{+}}$)}
 &\multicolumn{2}{c|}{${^2 P_{\frac{3}{2}}}$ (${J^P={\frac{3}{2}}^{+}}$)}
 &\multicolumn{2}{c|}{${^4 P_{\frac{3}{2}}}$ (${J^P={\frac{3}{2}}^{+}}$)}
 &\multicolumn{1}{c}{${^6 P_{\frac{3}{2}}}$ (${J^P={\frac{3}{2}}^{+}}$)}

  \\\cline{2-8}

$(Y, I, I_3)$  &$(0^{+}\otimes1^{+}\otimes
{\frac{1}{2}}^{-})_{\frac{1}{2}}\otimes1^{-}$
&$(0^{+}\otimes1^{+}\otimes
{\frac{1}{2}}^{-})_{\frac{3}{2}}\otimes1^{-}$
&$((1^{+}\otimes1^{+})_{0}\otimes{\frac{1}{2}}^{-})_{\frac{1}{2}}\otimes1^{-}$
         &$((1^{+}\otimes1^{+})_{1}\otimes{\frac{1}{2}}^{-})_{\frac{1}{2}}\otimes1^{-}$

               &$((1^{+}\otimes1^{+})_{1}\otimes{\frac{1}{2}}^{-})_{\frac{3}{2}}\otimes1^{-}$     &$((1^{+}\otimes1^{+})_{2}\otimes{\frac{1}{2}}^{-})_{\frac{3}{2}}\otimes1^{-}$
               &$((1^{+}\otimes1^{+})_{2}\otimes{\frac{1}{2}}^{-})_{\frac{5}{2}}\otimes1^{-}$\\
\hline

$(1,\frac{1}{2},\frac{1}{2})$  &2.836 & 2.111 &0.405
      &2.025 &1.218 &2.643 &1.903 \\

     \bottomrule[1pt]

 &\multicolumn{3}{c|}{${^{4}P_{\frac{5}{2}}^{+}}$ (${J^P={\frac{5}{2}}^{+}}$)}  &\multicolumn{1}{c|}{${^{6}P_{\frac{5}{2}}}$ (${J^P={\frac{5}{2}}^{+}}$)} &\multicolumn{1}{c|}{${^6P_{\frac{7}{2}}}$ (${J^P={\frac{7}{2}}^{+}}$)} \\\cline{2-8}

$(Y, I, I_3) $ &$(0^{+}\otimes1^{+}\otimes
{\frac{1}{2}}^{-})\otimes1^{-}$
         &$((1^{+}\otimes1^{+})_{1}\otimes{\frac{1}{2}}^{-})_{\frac{3}{2}}\otimes1^{-}$

        &$((1^{+}\otimes1^{+})_{2}\otimes{\frac{1}{2}}^{-})_{\frac{3}{2}}\otimes1^{-}$
               &$((1^{+}\otimes1^{+})_{2}\otimes{\frac{1}{2}}^{-})_{\frac{5}{2}}\otimes1^{-}$
               &$1^{+}\otimes1^{+}\otimes{\frac{1}{2}}^{-}\otimes1^{-}$\\
\hline

$(1,\frac{1}{2},\frac{1}{2})$ &3.245 &2.028 &3.972
      &2.748 &3.651\\

     \bottomrule[1pt] \bottomrule[1pt]
      \end{tabular}
  \end{center}
\end{table*}

\renewcommand{\arraystretch}{1.8}
\begin{table*}[htbp]
\tiny \caption{The magnetic moments of the pentaquark states in the
diquark-diquark-antiquark model with the configuration $(c
q_3)(q_1q_2)\bar c$ and the isospin
$(I,I_3)=(\frac{1}{2},\frac{1}{2})$. They are in the $8_{2f}$
representation from ${\bar3}_f\otimes 3_f=1_f\oplus8_{2f}$ and
$8_{1f}$ representation from ${6}_f\otimes 3_f=10_f\oplus8_{1f}$
respectively. The third line denotes the $J^P$ of the diquark
$(cq_3)$, diquark $(q_1q_2)$, antiquark and orbital excitation,
respectively. For the pentaquark states with positive parity, the
$\lambda$ mode P-wave excitation is between the $\bar c$ and center
of mass system of diquark $(c q_3)$ and diquark $(q_1q_2)$. The unit
is the magnetic moment of the proton.}\label{dda8l}
\begin{center}
   \begin{tabular}{c|c|c|c|c|c|c|cc} \toprule[1pt]
\multicolumn{8}{c} {Flavor representation-$8_{2f}$}\\\toprule[1pt]

 &\multicolumn{2}{c|}{$^2S_{\frac{1}{2}}$ ($J^P={\frac{1}{2}}^{-}$)}   &\multicolumn{1}{c|}{$^{4}S{{\frac{3}{2}}}$ (${J^P={\frac{3}{2}}^{-}}$)}  &\multicolumn{2}{c|}{${^{2}P_{\frac{1}{2}}}$ (${J^P={\frac{1}{2}}^{+}}$)}  &\multicolumn{1}{c|}{${^{4}P_{\frac{1}{2}}}$ (${J^P={\frac{1}{2}}^{+}}$)}\\\cline{2-7}
 \hline
$(Y, I, I_3)$  & ${0}^{+}\otimes0^{+}\otimes{\frac{1}{2}}^{-}
\otimes{0^{+}}$
      & ${1}^{+}\otimes0^{+}\otimes{\frac{1}{2}}^{-} \otimes{0^{+}}$  & ${1}^{+}\otimes0^{+}\otimes{\frac{1}{2}}^{-} \otimes{0^{+}}$  &  $0^{+}\otimes0^{+}\otimes{\frac{1}{2}}^{-} \otimes{1^{-}}$ &$({1}^{+}\otimes0^{+}\otimes{\frac{1}{2}}^{-})_{\frac{1}{2}} \otimes{1^{-}}$ &$({1}^{+}\otimes0^{+}\otimes{\frac{1}{2}}^{-})_{\frac{3}{2}}\otimes{1^{-}}$\\
\hline

$(1,\frac{1}{2},\frac{1}{2})$ &-0.403&1.667&1.895&0.675&0.854&0.348\\

\bottomrule[1pt]

&\multicolumn{2}{c|}{$^{2}P_{\frac{3}{2}}$
($J^P={\frac{3}{2}}^{+}$)}
&\multicolumn{1}{c|}{$^{4}P_{\frac{3}{2}}$
($J^P={\frac{3}{2}}^{+}$)}
&\multicolumn{1}{c|}{${^{4}P_{\frac{5}{2}}^{+}}$
(${J^P={\frac{5}{2}}^{+}}$)}   \\\cline{2-7} \hline
 $(Y, I, I_3)$  &
${0}^{+}\otimes0^{+}\otimes{\frac{1}{2}}^{-} \otimes{1^{-}}$
      & $({1}^{+}\otimes0^{+}\otimes{\frac{1}{2}}^{-} )_{\frac{1}{2}}\otimes{1^{-}}$ & $({1}^{+}\otimes0^{+}\otimes{\frac{1}{2}}^{-})_{\frac{3}{2}} \otimes{1^{-}}$ & ${1}^{+}\otimes0^{+}\otimes{\frac{1}{2}}^{-} \otimes{1^{-}}$ \\
\hline

$(1,\frac{1}{2},\frac{1}{2})$ &0.408&3.781&2.235 &4.009 \\

\bottomrule[1pt]
% \bottomrule[1pt]
 %     \end{tabular}
%  \end{center}
%\end{table*}

%\renewcommand{\arraystretch}{1.8}
%\begin{table*}[htbp]
%\caption{The magnetic moment of the pentaquark states in the
%diquark-diquark-antiquark model with the configuration $(c
%q_3)(q_1q_2)\bar c$ in the $8_{1f}$ flavor representation from
%${6}_f\otimes 3_f=10_f\oplus8_{1f}$. The third line denotes the
%$J^P$ of the diquark $(cq_3)$, diquark $(q_1q_2)$, anti-quark and
%orbital excitation respectively. For the pentaquark states with
%positive parity, the $\lambda$ mode P-wave excitation is between
%between the $\bar c$ and center of mass of diquark $(c q_3)$ and
%diquark $(q_1q_2)$. The unit is the magnetic moment of the proton.
%}\label{adda8-p1} \tiny
 % \begin{center}
%   \begin{tabular}{c|c|c|c|c|c|c|c} \toprule[1pt]\toprule[1pt]
\multicolumn{8}{c} {Flavor representation-$8_{1f}$}\\\toprule[1pt]
&\multicolumn{3}{c|}{$^{2}S_{\frac{1}{2}}$
($J^P={\frac{1}{2}}^{-}$)}
&\multicolumn{3}{c|}{${^{4}S_{\frac{3}{2}}}$
(${J^P={\frac{3}{2}}^{-}}$)}
&\multicolumn{1}{c}{$^{6}S_{\frac{5}{2}}$ ($J^P={\frac{5}{2}}^{-}$)}
\\\cline{2-8}

$(Y, I, I_3)$   & $0^{+}\otimes1^{+} \otimes
{\frac{1}{2}}^{-}\otimes0^{+}$
      &$(1^{+}\otimes1^{+})_{0} \otimes {\frac{1}{2}}^{-}\otimes0^{+}$  &$(1^{+}\otimes1^{+})_{1} \otimes {\frac{1}{2}}^{-}\otimes0^{+}$     &$(0^{+}\otimes1^{+})\otimes{\frac{1}{2}}^{-}\otimes0^{+}$ &$(1^{+}\otimes1^{+})_{1} \otimes {\frac{1}{2}}^{-}\otimes0^{+}$
       &$(1^{+}\otimes1^{+})_{2} \otimes {\frac{1}{2}}^{-}\otimes0^{+}$   &$(1^{+}\otimes1^{+})\otimes {\frac{1}{2}}^{-}\otimes0^{+}$
\\
\hline

$(1,\frac{1}{2},\frac{1}{2})$  &2.029 &-0.403 &1.216
      &2.439 &1.219 &3.163 &2.842 \\

    \bottomrule[1pt]

&\multicolumn{1}{c|}{$^{2}P_{\frac{1}{2}}$
(${J^P={\frac{1}{2}}^{+}}$)}
&\multicolumn{1}{c|}{$^{4}P_{\frac{1}{2}}$
(${J^P={\frac{1}{2}}^{+}}$)}
&\multicolumn{2}{c|}{$^{2}P_{\frac{1}{2}}$
(${J^P={\frac{1}{2}}^{+}}$)}
&\multicolumn{2}{c|}{$^{4}P_{\frac{1}{2}}$
(${J^P={\frac{1}{2}}^{+}}$)}\\\cline{2-8} $(Y, I, I_3)$
&$(0^{+}\otimes1^{+}\otimes
{\frac{1}{2}}^{-})_{\frac{1}{2}}\otimes1^{-}$
&$(0^{+}\otimes1^{+}\otimes
{\frac{1}{2}}^{-})_{\frac{3}{2}}\otimes1^{-}$
&$((1^{+}\otimes1^{+})_{0}\otimes{\frac{1}{2}}^{-})_{\frac{1}{2}}\otimes1^{-}$
         &$((1^{+}\otimes1^{+})_{1}\otimes{\frac{1}{2}}^{-})_{\frac{1}{2}}\otimes1^{-}$
 &$((1^{+}\otimes1^{+})_{1}\otimes{\frac{1}{2}}^{-})_{\frac{3}{2}}\otimes1^{-}$     &$((1^{+}\otimes1^{+})_{2}\otimes{\frac{1}{2}}^{-})_{\frac{3}{2}}\otimes1^{-}$\\
\hline

$(1,\frac{1}{2},\frac{1}{2})$ &0.441 &0.796 &1.237 &0.697 &0.126 &1.206 \\

     \bottomrule[1pt]

 &\multicolumn{1}{c|}{${^2 P_{\frac{3}{2}}}$ (${J^P={\frac{3}{2}}^{+}}$)}
 &\multicolumn{1}{c|}{${^4 P_{\frac{3}{2}}}$ (${J^P={\frac{3}{2}}^{+}}$)}
 &\multicolumn{2}{c|}{${^2 P_{\frac{3}{2}}}$ (${J^P={\frac{3}{2}}^{+}}$)}
 &\multicolumn{2}{c|}{${^4 P_{\frac{3}{2}}}$ (${J^P={\frac{3}{2}}^{+}}$)}
  &\multicolumn{1}{c}{${^6 P_{\frac{3}{2}}}$ (${J^P={\frac{3}{2}}^{+}}$)}
  \\\cline{2-8}

$(Y, I, I_3)$  &$(0^{+}\otimes1^{+}\otimes
{\frac{1}{2}}^{-})_{\frac{1}{2}}\otimes1^{-}$
&$(0^{+}\otimes1^{+}\otimes
{\frac{1}{2}}^{-})_{\frac{3}{2}}\otimes1^{-}$
&$((1^{+}\otimes1^{+})_{0}\otimes{\frac{1}{2}}^{-})_{\frac{1}{2}}\otimes1^{-}$
         &$((1^{+}\otimes1^{+})_{1}\otimes{\frac{1}{2}}^{-})_{\frac{1}{2}}\otimes1^{-}$

               &$((1^{+}\otimes1^{+})_{1}\otimes{\frac{1}{2}}^{-})_{\frac{3}{2}}\otimes1^{-}$     &$((1^{+}\otimes1^{+})_{2}\otimes{\frac{1}{2}}^{-})_{\frac{3}{2}}\otimes1^{-}$
               &$((1^{+}\otimes1^{+})_{2}\otimes{\frac{1}{2}}^{-})_{\frac{5}{2}}\otimes1^{-}$\\
\hline

$(1,\frac{1}{2},\frac{1}{2})$  &3.706 &2.459 &1.250 &2.870 &1.556 &2.981 &1.395\\

     \bottomrule[1pt]

 &\multicolumn{3}{c|}{${^{4}P_{\frac{5}{2}}^{+}}$ (${J^P={\frac{5}{2}}^{+}}$)}  &\multicolumn{1}{c|}{${^{6}P_{\frac{5}{2}}}$ (${J^P={\frac{5}{2}}^{+}}$)} &\multicolumn{1}{c|}{${^6P_{\frac{7}{2}}}$ (${J^P={\frac{7}{2}}^{+}}$)} \\\cline{2-8}

$(Y, I, I_3)$  &$(0^{+}\otimes1^{+}\otimes
{\frac{1}{2}}^{-})\otimes1^{-}$
         &$((1^{+}\otimes1^{+})_{1}\otimes{\frac{1}{2}}^{-})_{\frac{3}{2}}\otimes1^{-}$
        &$((1^{+}\otimes1^{+})_{2}\otimes{\frac{1}{2}}^{-})_{\frac{3}{2}}\otimes1^{-}$
               &$((1^{+}\otimes1^{+})_{2}\otimes{\frac{1}{2}}^{-})_{\frac{5}{2}}\otimes1^{-}$
               &$1^{+}\otimes1^{+}\otimes{\frac{1}{2}}^{-}\otimes1^{-}$\\
\hline

$(1,\frac{1}{2},\frac{1}{2})$ &4.116 &2.873 &4.817  &2.990 &4.496\\

     \bottomrule[1pt]\bottomrule[1pt]
      \end{tabular}
  \end{center}
\end{table*}

%{\color{blue}{In Tables \ref{dda8r} and \ref{dda8l}, the magnetic moments of
%the states with $J^P=\frac{1}{2}^-$ and configuration $0^{+}\otimes0^{+}\otimes\frac{1}{2}^-\otimes 0^+$ are all $-0.403\mu_p$. In this case,
%The reason is similar to that in the molecular scheme.
%the spin of the
%diquark $(cq_3)$ and diquark $(q_1q_2)$ are both $0$ in this
%configuration. Moreover, there is no orbital excitation. Thus, in
%this case, only the anti-charm quark contributes to the magnetic
%moment, which leads to the same negative magnetic moment even for the nine flavor configurations in Tables \ref{dda8+1} and \ref{dda8+1p1}.}}
%for the nine flavor configurations.
%The third column of the two tables
%shares the same pattern as those in Table \ref{8+1} due to the
%vanishing contributions from the diquark $q_1q_2$ and orbital
%excitation.

The numerical results for the $\frac{5}{2}^{-}$ pentaquark states in
$8_{1f}$ representation are the same as those in the molecular model
and diquark-diquark-antiquark model, which is illustrated in Tables
\ref{pcm}, \ref{dda8r}, and \ref{dda8l}. The magnetic moments of the
$\frac{5}{2}^{-}$ pentaquark states in the diquark-triquark scheme
are also the same as those in the above two models, which can be
seen in Sec. \ref{sec5}. For the $\frac{5}{2}^{-}$ pentaquark
states, the analytic formulas for the magnetic moments are the same
in different models as illustrated in Tables \ref{analysis3} and
\ref{analysis8+1}. The sum of the magnetic moments of all the
constituent quarks leads to those of the pentaquark states.
Therefore, the magnetic moments of the $\frac{5}{2}^{-}$ only depend
on the flavor content and do not dependent on their configurations
and the models.

The $P_c(4380)$ and $P_c(4450)$ with the isospin
$(I,I_3)=(\frac{1}{2},\frac{1}{2})$ are in the $8_{1f}$ or $8_{2f}$
representation in the diquark-diquark-antiquark scheme. The
numerical results for the states with the $\rho$ and $\lambda$
excitation modes are different. The magnetic moments of the $P_c$
states in the $8_{1f}$ representation with $J^P=\frac{3}{2}^{+},
\frac{5}{2}^{+}$ and $\lambda$ excitation are larger than those with
the $\rho$ excitation except the configuration
${((1^+\otimes1^+)_2\otimes \frac{1}{2} })_{\frac{5}{2}}\otimes1^-$,
as illustrated in Tables \ref{dda8r} and \ref{dda8l}.

%%%%%%%%%%%%%%%%%%%%%%%%%%%%%%
\section{Magnetic moment of the $P_c$ states in the diquark-triquark model}
\label{sec5}
%%%%%%%%%%%%%%%%%%%%%%%%%%%%%%%

The diquark-triquark configurations are either $(c q_3)(\bar c
q_1q_2)$ or $(q_1q_2)(\bar c cq_3)$. Both the $(c q_3)$ diquark
within $(c q_3)(\bar c q_1 q_2)$ and $(q_1q_2)$ diquark within
$(q_1q_2)(\bar c cq_3)$ are in a ${\bar 3}_c$ representation. The
triquark $\bar c q_1q_2$ or ($\bar c cq_3$) is also in the $3_c$
representation.

However, as pointed out in Sec \ref{sec2}, the diquark $q_1q_2$ in
the triquark $(\bar c q_1 q_2)$ can be either in a ${\bar 3}_c$ or
$6_c$ representation. When the $q_1q_2$ is in the ${\bar 3}_c$
representation, the structures are the same as that in Table
\ref{c1}. Their wave functions are listed in Table
\ref{dt10-flavor}. When the $q_1q_2$ is in the $6_c$ representation,
the structures are listed in Table \ref{c3}.
\renewcommand{\arraystretch}{1.8}
\begin{table*}[htbp]
\caption{ The flavor wave functions of the pentaquark states in the
diquark-triquark model with the configuration $(c q_3)(q_1q_2\bar
c)$ in different flavor representations from ${3}_f\otimes 3_f\otimes
3_f=10_f\oplus8_{1f}\oplus8_{2f}\oplus1_f$. Here,
$\{q_1q_2\}=\frac{1}{\sqrt 2}(q_1q_2+q_2q_1)$ and
$[q_1q_2]=\frac{1}{\sqrt 2}(q_1q_2-q_2q_1)$.}\label{dt10-flavor}
\begin{center}
\begin{tabular}{c|c|c} \toprule[1pt]\toprule[1pt]

$(Y, I, I_3)$ & Wave function-$8_{1f}$   & Wave function-$8_{2f}$\\
\midrule[1pt] $(1,\frac{1}{2},\frac{1}{2})$ &$-\sqrt{\frac{1}{3}}({
c}u)(\bar c\{ud\})+\sqrt{\frac{2}{3}}({c}d)(\bar c\{uu\})$
& $({c}u)(\bar c[ud])$\\

$(1,\frac{1}{2},-\frac{1}{2})$  &$\sqrt{\frac{1}{3}}({c}d)(\bar
c\{ud\})-\sqrt{\frac{2}{3}}({c}u)(\bar c\{dd\})$
&  $({c}d)(\bar c[ud])$\\

$(-1,\frac{1}{2},\frac{1}{2})$ &$\sqrt{\frac{1}{3}}({c}s)(\bar
c\{us\})-\sqrt{\frac{2}{3}}({c}u)(\bar c\{ss\})$
&  $({c}s)(\bar c[us])$ \\

$(-1,\frac{1}{2},-\frac{1}{2})$  &$\sqrt{\frac{1}{3}}({c}s)(\bar
c\{ds\})-\sqrt{\frac{2}{3}}({c}d)(\bar c\{ss\})$
&        $({c}s)(\bar c[ds])$\\

$ (0,1,1)$& $\sqrt{\frac{1}{3}}({c}u)(\bar
c\{us\})-\sqrt{\frac{2}{3}}({c}s)(\bar c\{uu\})$
&      $({c}u)(\bar c[us])$ \\

$ (0,1,0)$& $\sqrt{\frac{1}{6}}[({c}d)(\bar c\{us\})+({c}u)(\bar
c\{ds\})]-\sqrt{\frac{2}{3}}({c}s)(\bar c\{ud\})$
& $\frac{1}{\sqrt2}\{ ({c}d)(\bar c[us])+({c}u)(\bar c[ds]) \}$\\

$ (0,1,-1)$& $\sqrt{\frac{1}{3}}({c}d)(\bar
c\{ds\})-\sqrt{\frac{2}{3}}({c}s)(\bar c\{dd\})$
& $({c}d)(\bar c[ds])$ \\

$ (0,0,0)$& $\sqrt{\frac{1}{2}}[({c}u)(\bar c\{ds\})-({c}d)(\bar
c\{us\})]$
&    $\frac{1}{\sqrt6} \{ ({c}d)(\bar c[us])-({ c}u)(\bar c[ds])-2( c s)( \bar c[ud]) \}$\\

     \bottomrule[1pt]

$(Y, I, I_3)$ &\multicolumn{1}{c|} {Wave function-$10_f$}
&Singlet\\
\midrule[1pt] $(1,\frac{3}{2},\frac{3}{2})$& $({c}u)(\bar c\{uu\})$
&   $\frac{1}{\sqrt3} \{ ({c}d)(\bar c[us])-({c}u)(\bar c[ds])+(c s)(\bar c[ud]) \}$\\

$(1,\frac{3}{2},\frac{1}{2})$  &  $\sqrt{\frac{2}{3}}({c}u)(\bar c\{ud\})+\sqrt{\frac{1}{3}}({c}d)(\bar c\{uu\})$ \\
$(1,\frac{3}{2},-\frac{1}{2})$&  $\sqrt{\frac{2}{3}}({c}d)(\bar c\{ud\})+\sqrt{\frac{1}{3}}({c}u)(\bar c\{dd\})$  \\
$(1,\frac{3}{2},-\frac{3}{2})$&  $({c}d)(\bar c\{dd\})$\\
$ (0,1,1)$& $\sqrt{\frac{2}{3}}({c}u)(\bar c\{us\})+\sqrt{\frac{1}{3}}({c}s)(\bar c\{uu\})$ \\
$ (0,1,0)$& $\sqrt{\frac{1}{3}}[({c}d)(\bar c\{us\})+({c}u)(\bar c\{ds\})]+\sqrt{\frac{1}{3}}({c}s)(\bar c\{ud\})$ \\
$ (0,1,-1)$& $\sqrt{\frac{2}{3}}({c}d)(\bar c\{ds\})+\sqrt{\frac{1}{3}}({c}s)(\bar c\{dd\})$\\
$ (-1,\frac{1}{2},\frac{1}{2})$& $\sqrt{\frac{1}{3}}({c}u)(\bar c\{ss\})+\sqrt{\frac{2}{3}}({c}s)(\bar c\{us\})$ \\
$ (-1,\frac{1}{2},-\frac{1}{2})$& $\sqrt{\frac{1}{3}}({c}d)(\bar c\{ss\})+\sqrt{\frac{2}{3}}({c}s)(\bar c\{ds\})$ \\
$(-2,0,0)$&  $({c}s)(\bar c\{ss\})$ \\
     \bottomrule[1pt]\bottomrule[1pt]

 \end{tabular}
  \end{center}
\end{table*}

\renewcommand{\arraystretch}{1.8}
\begin{table}[htbp]
\caption{Structure of the $6_c$ $q_1q_2$
diquark within the triquark in the diquark-triquark model. Here, S and A represent the wave function of the corresponding space are symmetry and antisymmetry, respectively. %$J_{baryon}=\frac{1}{2}$.
}\label{c3}
\begin{tabular}{c|ccc|c|cccc}
 \toprule[1pt]
\multirow{4}{*}{$J_{baryon}=\frac{1}{2}$}& Color & ${6}_c$& S &\multirow{4}{*}{$J_{baryon}=\frac{1}{2},\frac{3}{2}$}& Color & ${6}_c$& S \\
%\hline
 & Space  & $L=0$ &  S&& Space  & $L=0$ &  S  \\
%\hline
 &Spin  & $s=0$ &  A & &Spin  & $s=1$ &  S \\
%\hline
 &Flavor  & ${6}_f$ &  S &&Flavor  & ${\bar3}_f$ &  A \\
\bottomrule[1pt]
\end{tabular}
\end{table}

The diquark-triquark model has a similar configuration as the
molecular model except the color representation. We also assume that
the P-wave excitation exists between the two clusters. The
derivation of the magnetic moment in this model is similar to that
of the molecule case. Now we have
\begin{eqnarray} \label{w7}
 \vec{\mu}&=&g_{d}{\mu_{d}}{\vec{S}_d}+g_{t}{\mu_t}{\vec{S}}_t+{\mu_l}\vec{l} \nonumber\\
g_{d}{\mu_{d}}{\vec{S}_d}&=&g_{Q} {\mu_{Q}}\vec{\frac{1}{2}}+g_{q_3}{\mu_{q_3}}\vec{\frac{1}{2}} \nonumber\\
 g_{t}{\mu_t}{\vec{S}}_t&=&g_{\bar Q} {\mu_{\bar Q}}\vec{\frac{1}{2}}+g_{q_1}{\mu_{q_1}}\vec{\frac{1}{2}}+g_{q_2}{\mu_{q_2}}\vec{\frac{1}{2}}\nonumber\\
{\mu_l}&=&\frac{{m_t}\mu_{d}+{m_d}\mu_{t}}{m_d+m_t}.
\end{eqnarray}
The subscripts $d$ and $t$ represent the diquark and triquark,
respectively. Similarly,
\begin{eqnarray} \label{w55}
|JJ\rangle=\left|\left[\left[ \left((q_1q_2)_{s_{12}}{\bar
c}_{\frac{1}{2}}\right)_{S_t}\otimes ( cq_3)_{S_d}\right]_S\otimes
l\right]_{J}^{J}\right\rangle,
\end{eqnarray}
\begin{eqnarray} \label{w6}
\mu &=&\langle JJ|g_{d}{\mu_{d}}{\vec{S}_d}+g_{t}{\mu_t}{\vec{S}}_t+{\mu_l}\vec{l}|JJ\rangle \nonumber\\
&&=\sum_{m, {S_z}}\langle lm,SS_z|JJ\rangle^2 \left \{
\mu_{l}m+\sum_{S'_z}
\left\langle {S_t}S'_z,S_d (S_z-S'_z)|SS_z\right\rangle^2 \left[ (S_z-S'_z)(\mu_{ Q}+\mu_{q_3})\right.\right.\nonumber\\
&&\left.\left. +\left\langle
{\frac{1}{2}}{\frac{1}{2}},s_{12}\left(S'_z-\frac{1}{2}\right)\bigg|S_t
S'_z\right\rangle^2\left(\mu_{\bar
Q}+\left(S'_z-\frac{1}{2}\right)(\mu_{q_1}+\mu_{q_2})\right)+\left\langle
{\frac{1}{2}}{-\frac{1}{2}},s_{12}\left(S'_z+\frac{1}{2}\right)\bigg|S_t
S'_z\right\rangle^2\left(-\mu_{\bar
Q}+\left(S'_z+\frac{1}{2}\right)(\mu_{q_1}+\mu_{q_2})\right)\right] \right\}.\nonumber\\
\end{eqnarray}
where we have replaced the spin of the baryon $S_b$ and the spin of
the meson $S_m$ in Eq. (\ref{w2}) and Eq. (\ref{w3}) by the spin of
the diquark $S_d$ and the spin of the triquark $S_t$. $m$, $S_z$,
and $S'_z$ are the third component of the orbital excitation, the
pentaquark total spin and the triquark spin respectively. $s_{12}$
is the total spin of the diquark inside the triquark.

The masses of the triquark are roughly the sum of the mass of the
corresponding diquark and the antiquark. When the $q_1q_2$ is in the
$6_c$ representation, their masses are not known. Therefore, we omit
results in this case. When the diquark is in the $\bar 3_c$
representation, the numerical results for the pentaquark states with
isospin $\frac{1}{2}$ are in Table \ref{dt}. We also list the
analytic formulae of the magnetic moments of the pentaquarks with
differen spin and configurations in Table \ref{analysis8+1}. For the
pentaquarks with the configuration $(q_1q_2)(\bar c cq_3)$, we
collect the flavor wave functions in Table \ref{ccbardt10-flavor},
and the numerical results for the
$(I,I_3)=(\frac{1}{2},\frac{1}{2})$ states are listed in Table
\ref{adanti8+1ccbar}.

\renewcommand{\arraystretch}{1.8}
\begin{table*}[htbp]
\caption{The magnetic moment of the pentaquark states as a
diquark-triquark state $(c q_3)(q_1q_2\bar c)$ with
$(I,I_3)=(\frac{1}{2},\frac{1}{2})$. The $8_{1f}$ and $8_{2f}$
flavor representations are from the
${\bar3}_f\otimes3_f=1_f\oplus8_{2f}$ and ${6}_f\otimes
 3_f=10_f\oplus8_{1f}$, respectively. The wave functions of the pentaquark states are listed in Table
\ref{dt10-flavor}. The third line denotes the $J^P$ of the triquark
$q_1q_2\bar c$, the diquark $c q_3$, and the orbital excitation. The unit
is the magnetic moments of the proton.}\label{dt}
\begin{center}
   \begin{tabular}{c|c|c|c|c|c|c|c} \toprule[1pt]\toprule[1pt]
\multicolumn{8}{c} {Flavor representation-$8_{2f}$}\\\toprule[1pt]
&\multicolumn{2}{c|}{$^2S_{\frac{1}{2}}$ ($J^P={\frac{1}{2}}^{-}$)}
&\multicolumn{1}{c|}{$^{4}S{{\frac{3}{2}}}$
(${J^P={\frac{3}{2}}^{-}}$)}
&\multicolumn{2}{c|}{${^{2}P_{\frac{1}{2}}}$
(${J^P={\frac{1}{2}}^{+}}$)}
&\multicolumn{1}{c|}{${^{4}P_{\frac{1}{2}}}$
(${J^P={\frac{1}{2}}^{+}}$)}\\\cline{2-7} \hline
 $(Y, I, I_3)$    &
${\frac{1}{2}}^{-}\otimes0^{+}\otimes0^{+}$
      & ${\frac{1}{2}}^{-}\otimes1^{+}\otimes0^{+}$   & ${\frac{1}{2}}^{-}\otimes1^{+}\otimes0^{+}$  &  ${\frac{1}{2}}^{-}\otimes0^{+}\otimes1^{-}$ &$({\frac{1}{2}}^{-}\otimes1^{+})_{\frac{1}{2}}\otimes1^{-}$
      & $({\frac{1}{2}}^{-}\otimes1^{+})_{\frac{3}{2}}\otimes1^{-}$ \\
\hline

$(1,\frac{1}{2},\frac{1}{2})$&-0.403&1.667&1.895&0.317&-0.384&0.967\\

\midrule[1pt]

 &\multicolumn{2}{c|}{$^{2}P_{\frac{3}{2}}$ ($J^P={\frac{3}{2}}^{+}$)} &\multicolumn{1}{c|}{$^{4}P_{\frac{3}{2}}$ ($J^P={\frac{3}{2}}^{+}$)}   &\multicolumn{1}{c|}{${^{4}P_{\frac{5}{2}}^{+}}$ (${J^P={\frac{5}{2}}^{+}}$)}   \\\cline{2-7}
\hline $(Y, I, I_3)$ & ${\frac{1}{2}}^{-}\otimes 0^{+}\otimes1^{-}$
      & $({\frac{1}{2}}^{-}\otimes1^{+})_{\frac{1}{2}}\otimes1^{-}$  &  $({\frac{1}{2}}^{-}\otimes1^{+})_{\frac{3}{2}}\otimes1^{-}$ & ${\frac{1}{2}}^{-}\otimes1^{+}\otimes1^{-}$ \\
\hline

$(1,\frac{1}{2},\frac{1}{2})$ &-0.129&1.924&1.493&2.153\\

     \bottomrule[1pt]
     %\bottomrule[1pt]
     %  \end{tabular}
   %\end{center}
 %\end{table*}

 %\renewcommand{\arraystretch}{1.8}
 %\begin{table*}[htbp]
 %\caption{The magnetic moment of the pentaquark states in the
 %diquark-triquark  model with the configuration $(c q_3)(q_1q_2\bar
 %c)$ in the $8_{1f}$ flavor representation from ${6}_f\otimes
 %3_f=10_f\oplus8_{1f}$. The third line denotes the $J^P$ of the
 %triquark $q_1q_2\bar c$, the diquark $c q_3$ and orbital excitation.
 %The unit is the magnetic moment of the proton.}\label{adt8-}
    %\scriptsize
 %\begin{center}

 % \begin{tabular}{c|c|c|c|c|c|c|c} \toprule[1pt]\toprule[1pt]
\multicolumn{8}{c} {Flavor representation-$8_{1f}$}\\\toprule[1pt]

  &\multicolumn{3}{c|}{$^{2}S_{\frac{1}{2}}$ ($J^P={\frac{1}{2}}^{-}$)}   &\multicolumn{3}{c|}{${^{4}S_{\frac{3}{2}}}$ (${J^P={\frac{3}{2}}^{-}}$)} &\multicolumn{1}{c}{$^{6}S_{\frac{5}{2}}^{-}$ ($J^P={\frac{5}{2}}^{-}$)} \\\cline{2-8}
$(Y, I, I_3)$  & ${\frac{1}{2}}^{-}\otimes0^{+}\otimes0^{+}$
      & ${\frac{1}{2}}^{-}\otimes1^{+}\otimes0^{+}$  &  ${\frac{3}{2}}^{-}\otimes1^{+}\otimes0^{+}$      & ${\frac{1}{2}}^{-}\otimes1^{+}\otimes0^{+}$ &${\frac{3}{2}}^{-}\otimes0^{+}\otimes0^{+}$
      & ${\frac{3}{2}}^{-}\otimes1^{+}\otimes0^{+}$  & ${\frac{3}{2}}^{-}\otimes1^{+}\otimes0^{+}$
\\
\hline

$(1,\frac{1}{2},\frac{1}{2})$ & 2.029 & - 0.408 & 1.221 & 2.433 & 2.439 & 1.950 & 2.842\\

     \bottomrule[1pt]

 &\multicolumn{2}{c|}{${^{2}p_{\frac{1}{2}}}$ (${J^P={\frac{1}{2}}^{+}}$)} &\multicolumn{2}{c|}{${^{4}p_{\frac{1}{2}}}$ (${J^P={\frac{1}{2}}^{+}}$)} &\multicolumn{1}{c|}{${^{2}p_{\frac{1}{2}}}$ (${J^P={\frac{1}{2}}^{+}}$)} &\multicolumn{1}{c|}{${^{4}p_{\frac{1}{2}}}$ (${J^P={\frac{1}{2}}^{+}}$)}  \\\cline{2-8}
$(Y, I, I_3)$      & ${\frac{1}{2}}^{-}\otimes0^{+}\otimes1^{-}$  &  $[{\frac{1}{2}}^{-}\otimes1^{+}]_{\frac{1}{2}}\otimes1^{-}$      & $[{\frac{1}{2}}^{-}\otimes1^{+}]_{\frac{3}{2}}\otimes1^{-}$ &${\frac{3}{2}}^{-}\otimes0^{+}\otimes1^{-}$    & $[{\frac{3}{2}}^{-}\otimes1^{+}]_{\frac{1}{2}}\otimes1^{-}$   & $[{\frac{3}{2}}^{-}\otimes1^{+}]_{\frac{3}{2}}\otimes1^{-}$\\
\hline

$(1,\frac{1}{2},\frac{1}{2})$  &-0.522 &0.286 &1.276&1.278&-0.256&1.008 \\

     \bottomrule[1pt]

  &\multicolumn{2}{c|}{${^{2}P_{\frac{3}{2}}}$ (${J^P={\frac{3}{2}}^{+}}$)} &\multicolumn{2}{c|}{${^{4}P_{\frac{3}{2}}}$ (${J^P={\frac{3}{2}}^{+}}$)}  &\multicolumn{1}{c|}{${^{2}P_{\frac{3}{2}}}$ (${J^P={\frac{3}{2}}^{+}}$)}  &\multicolumn{1}{c|}{${^{4}P_{\frac{3}{2}}}$ (${J^P={\frac{3}{2}}^{+}}$)}  &\multicolumn{1}{c}{${^{6}P_{\frac{3}{2}}}$ (${J^P={\frac{3}{2}}^{+}}$)}  \\\cline{2-8}
$ (Y, I, I_3)$      & ${\frac{1}{2}}^{-}\otimes0^{+}\otimes1^{-}$  &  $[{\frac{1}{2}}^{-}\otimes1^{+}]_{\frac{1}{2}}\otimes1^{-}$      & $[{\frac{1}{2}}^{-}\otimes1^{+}]_{\frac{3}{2}}\otimes1^{-}$ &${\frac{3}{2}}^{-}\otimes0^{+}\otimes1^{-}$    & $[{\frac{3}{2}}^{-}\otimes1^{+}]_{\frac{1}{2}}\otimes1^{-}$   & $[{\frac{3}{2}}^{-}\otimes1^{+}]_{\frac{3}{2}}\otimes1^{-}$   & $[{\frac{3}{2}}^{-}\otimes1^{+}]_{\frac{5}{2}}\otimes1^{-}$\\
\hline

$(1,\frac{1}{2},\frac{1}{2})$ &2.262&-0.182&1.874&1.882&1.446&1.520&2.252\\

     \bottomrule[1pt]

 &\multicolumn{3}{c|}{${^{4}P_{\frac{5}{2}}}$ (${J^P={\frac{5}{2}}^{+}}$)} &\multicolumn{1}{c|}{${^{6}P_{\frac{5}{2}}}$ (${J^P={\frac{5}{2}}^{+}}$)}  &\multicolumn{1}{c|}{${^{6}P_{\frac{7}{2}}}$ (${J^P={\frac{7}{2}}^{+}}$)}\\\cline{2-8}
$ (Y, I, I_3)$&  ${\frac{1}{2}}^{-}\otimes1^{+}\otimes1^{-}$ &${\frac{3}{2}}^{-}\otimes0^{+}\otimes1^{-}$    & $[{\frac{3}{2}}^{-}\otimes1^{+}]_{\frac{3}{2}}\otimes1^{-}$   & $[{\frac{3}{2}}^{-}\otimes1^{+}]_{\frac{5}{2}}\otimes1^{-}$   & ${\frac{3}{2}}^{-}\otimes1^{+}\otimes1^{-}$\\
\hline

$(1,\frac{1}{2},\frac{1}{2})$ &2.658&2.671&2.176&2.582&3.068 \\

     \bottomrule[1pt] \bottomrule[1pt]
      \end{tabular}
  \end{center}
\end{table*}

\renewcommand{\arraystretch}{1.6}
\begin{table*}[htbp]
\caption{The analytic formula for the magnetic moment of the
pentaquark state in the configuration $(Q q_3)(\bar Q q_1q_2)$ when
$s_{q_1q_2}=0$ or $1$.}\label{analysis8+1}
% \scriptsize
\begin{center}
   \begin{tabular}{c|c|c|c} \toprule[1pt]\toprule[1pt]
\multicolumn{4}{c} {$s_{q_1q_2}=0$ }\\\toprule[1pt] $J^P$
&$^{2s+1}L_J$ & &Formula \\\midrule[1pt]

{$\frac{1}{2}^{-}$}  &{{${^{2}S_{\frac{1}{2}}}$}}   &${\frac{1}{2}}^{-}\otimes0^{+}\otimes0^{+}$ &$\mu _{\bar{Q}}$\\

&&${\frac{1}{2}}^{-}\otimes1^{+}\otimes0^{+}$  &$\frac{1}{3} \left(2 \mu _Q+2 \mu _{{q_3}}-\mu _{\bar{Q}}\right)$\\
\hline

{$\frac{3}{2}^{-}$}  &{{${^{4}S_{\frac{3}{2}}}$}} & ${\frac{1}{2}}^{-}\otimes1^{+}\otimes0^{+}$ &$\mu _Q+\mu _{{q_3}}+\mu _{\bar{Q}}$\\

\hline

{$\frac{1}{2}^{+}$}  &{{${^{2}P_{\frac{1}{2}}}$}} & ${\frac{1}{2}}^{-}\otimes0^{+}\otimes1^{-}$ &$\frac{1}{3} \left(2 \mu _1-\mu _{\bar{Q}}\right)$\\

 &  &  $({\frac{1}{2}}^{-}\otimes1^{+})_{\frac{1}{2}}\otimes1^{-}$ &$\frac{1}{9} \left(6 \mu _1-2 \mu _Q-2 \mu_{{q_3}}+\mu _{\bar{Q}}\right)$
 \\\cline{2-4}

&\multirow{1}{*}{${^{4}P_{\frac{1}{2}}}$} &
$({\frac{1}{2}}^{-}\otimes1^{+})_{\frac{3}{2}}\otimes1^{-}$
 &$\frac{1}{9}(-3\mu_1+5 (\mu_Q+\mu _{q_3}+\mu_{\bar{Q}})$\\
 \hline

 {$\frac{3}{2}^{+}$}  &{{${^{2}P_{\frac{3}{2}}}$}} & ${\frac{1}{2}}^{-}\otimes 0^{+}\otimes1^{-}$ &$\mu _1+\mu_{\bar{Q}}$\\

    & &$({\frac{1}{2}}^{-}\otimes1^{+})_{\frac{1}{2}}\otimes1^{-}$ &$\frac{1}{3} \left(3 \mu _1+2 \mu _Q+2 \mu _{{q_3}}-\mu _{\bar{Q}}\right)$\\\cline{2-4}

    &\multirow{1}{*}{${^{4}P_{\frac{3}{2}}}$} &  $({\frac{1}{2}}^{-}\otimes1^{+})_{\frac{3}{2}}\otimes1^{-}$ &$\frac{1}{15} (6 \mu _1+11(\mu _Q+\mu _{{q_3}}+\mu _{\bar{Q}}))$\\
\hline

   \multirow{1}{*}{$\frac{5}{2}^{+}$}  &{\multirow{1}{*}{${^{4}P_{\frac{5}{2}}}$}} & ${\frac{1}{2}}^{-}\otimes1^{+}\otimes1^{-}$  &$\mu _1+\mu _Q+\mu _{{q_3}}+\mu _{\bar{Q}}$\\

\bottomrule[1pt]
%\bottomrule[1pt]

   %   \end{tabular}
  %\end{center}
%\end{table*}

%\renewcommand{\arraystretch}{1.8}
%\begin{table*}[htbp]
%\caption{The analytic formula for the magnetic moment of the
%pentaquark state in the configuration $(Q q_3)(\bar Q q_1q_2)$ when
%$s_{q_1q_2}=1$.}\label{ananlysis 210-}
   %\scriptsize
%\begin{center}
%   \begin{tabular}{c|c|c|c} \toprule[1pt]\toprule[1pt]
\multicolumn{4}{c} {$s_{q_1q_2}=1$ }\\\toprule[1pt]
\,\,\,\,$J^P$\,\,\,\,    &\,$^{2s+1}L_J$\,  &   &Formula
\\\midrule[1pt]

{$\frac{1}{2}^{-}$}  &{{${^{2}S_{\frac{1}{2}}}$}}   & ${\frac{1}{2}}^{-}\otimes0^{+}\otimes0^{+}$ &$\frac{1}{3} (2 \mu _{{q_1}}+2 \mu _{{q_2}}-\mu _{\bar{Q}})$\\

&&  ${\frac{1}{2}}^{-}\otimes1^{+}\otimes0^{+}$ &$\frac{1}{9}(6 \mu _Q-2 \mu _{{q_1}}-2 \mu _{{q_2}}+6\mu_{{q_3}}+\mu _{\bar{Q}})$\\

&&${\frac{3}{2}}^{-}\otimes1^{+}\otimes0^{+}$ &$\frac{1}{9}(-3\mu _Q+5 \mu _{{q_1}}+5 \mu _{{q_2}}-3 \mu _{{q_3}}+5 \mu _{\bar{Q}})$\\
\hline

{$\frac{3}{2}^{-}$} &{{${^{4}S_{\frac{3}{2}}}$}}
 &${\frac{1}{2}}^{-}\otimes 0^{+}\otimes0^{+}$ &$\mu _Q+\frac{2\mu _{{q_1}}}{3}+\frac{2\mu _{q_2}}{3}+\mu _{{q_3}}-\frac{\mu _{\bar{Q}}}{3}$\\

&&${\frac{3}{2}}^{-}\otimes0^{+}\otimes0^{+}$ &$\mu _{{q_1}}+\mu _{{q_2}}+\mu _{\bar{Q}}$\\

&&${\frac{3}{2}}^{-}\otimes1^{+}\otimes0^{+}$ &$\frac{1}{15} (6 \mu _Q+11 \mu _{{q_1}}+11 \mu _{{q_2}}+6 \mu _{{q_3}}+11 \mu _{\bar{Q}})$\\

\hline

{$\frac{5}{2}^{-}$}  &{{${^{6}S_{\frac{5}{2}}}$}}   &${\frac{3}{2}}^{-}\otimes1^{+}\otimes0^{+}$ &$\mu _Q+\mu _{{q_1}}+\mu _{{q_2}}+\mu _{{q_3}}+\mu _{\bar{Q}}$\\

\hline {$\frac{1}{2}^{+}$}
&{{${^{2}P_{\frac{1}{2}}}$}}   &${\frac{1}{2}}^{-}\otimes0^{+}\otimes1^{-}$ &$\frac{1}{9} \left(6 \mu _1-2 \mu _{{q_1}}-2 \mu _{{q_2}}+\mu _{\bar{Q}}\right)$\\

&  &$[{\frac{1}{2}}^{-}\otimes1^{+}]_{\frac{1}{2}}\otimes1^{-}$
&$\frac{1}{27} \left(18 \mu _1-6 \mu _Q+2 \mu _{{q_1}}+2 \mu
_{{q_2}}-6 \mu _{{q_3}}-\mu _{\bar{Q}}\right)$\\\cline{2-4}

&{{${^{4}P_{\frac{1}{2}}}$}}   & $[{\frac{1}{2}}^{-}\otimes1^{+}]_{\frac{3}{2}}\otimes1^{-}$ &$\frac{1}{27} \left(-9 \mu _1+5 \left(3 \mu _Q+2 \mu _{{q_1}}+2 \mu _{{q_2}}+3 \mu _{{q_3}}-\mu _{\bar{Q}}\right)\right)$\\

&  &${\frac{3}{2}}^{-}\otimes0^{+}\otimes1^{-}$  &$\frac{1}{9}
\left(-3 \mu _1+5 \left(\mu _{{q_1}}+\mu _{{q_2}}+\mu
_{\bar{Q}}\right)\right)$\\\cline{2-4}

&{\multirow{1}*{${^{2}P_{\frac{1}{2}}}$}}   & $[{\frac{3}{2}}^{-}\otimes1^{+}]_{\frac{1}{2}}\otimes1^{-}$ &$\frac{1}{27} \left(18 \mu _1+3 \mu _Q-5 \mu _{{q_1}}-5 \mu _{{q_2}}+3 \mu _{{q_3}}-5 \mu _{\bar{Q}}\right)$\\

&{\multirow{1}*{${^{4}P_{\frac{1}{2}}}$}}   &  $[{\frac{3}{2}}^{-}\otimes1^{+}]_{\frac{3}{2}}\otimes1^{-}$ &$\frac{1}{27} \left(-9 \mu _1+6 \mu _Q+11 \mu _{{q_1}}+11 \mu _{{q_2}}+6 \mu _{{q_3}}+11 \mu _{\bar{Q}}\right)$\\

\hline
{$\frac{3}{2}^{+}$}  &{{${^{2}P_{\frac{3}{2}}}$}}   &${\frac{1}{2}}^{-}\otimes0^{+}\otimes1^{-}$ &$\frac{1}{3} \left(3 \mu _1+2 \mu _{{q_1}}+2 \mu _{{q_2}}-\mu _{\bar{Q}}\right)$\\

&  &$[{\frac{1}{2}}^{-}\otimes1^{+}]_{\frac{1}{2}}\otimes1^{-}$
&$\frac{1}{9} \left(9 \mu _1+6 \mu _Q-2 \mu _{{q_1}}-2 \mu
_{{q_2}}+6 \mu _{{q_3}}+\mu _{\bar{Q}}\right)$\\\cline{2-4}

&{{${^{4}P_{\frac{3}{2}}}$}}   & $[{\frac{1}{2}}^{-}\otimes1^{+}]_{\frac{3}{2}}\otimes1^{-}$ &$\frac{1}{45} \left(18 \mu _1+11 \left(3 \mu _Q+2 \mu _{{q_1}}+2 \mu _{{q_2}}+3 \mu _{{q_3}}-\mu _{\bar{Q}}\right)\right)$\\

&  &${\frac{3}{2}}^{-}\otimes0^{+}\otimes1^{-}$  &$\frac{1}{15}
\left(6 \mu _1+11 \left(\mu _{{q_1}}+\mu _{{q_2}}+\mu
_{\bar{Q}}\right)\right)$\\\cline{2-4}

&{\multirow{1}*{${^{2}P_{\frac{3}{2}}}$}}   & $[{\frac{3}{2}}^{-}\otimes1^{+}]_{\frac{1}{2}}\otimes1^{-}$ &$\frac{1}{9} \left(9 \mu _1-3 \mu _Q+5 \mu _{{q_1}}+5 \mu _{{q_2}}-3 \mu _{{q_3}}+5 \mu _{\bar{Q}}\right)$\\

&{\multirow{1}*{${^{4}P_{\frac{3}{2}}}$}}   &  $[{\frac{3}{2}}^{-}\otimes1^{+}]_{\frac{3}{2}}\otimes1^{-}$ &$\frac{1}{225} \left(90 \mu _1+11 \left(6 \mu _Q+11 \mu _{{q_1}}+11 \mu _{{q_2}}+6 \mu _{{q_3}}+11 \mu _{\bar{Q}}\right)\right)$\\

&{\multirow{1}*{${^{6}P_{\frac{3}{2}}}$}}   &  $[{\frac{3}{2}}^{-}\otimes1^{+}]_{\frac{5}{2}}\otimes1^{-}$ &$\frac{3}{25} \left(-5 \mu _1+7 \left(\mu _Q+\mu _{{q_1}}+\mu _{{q_2}}+\mu _{{q_3}}+\mu _{\bar{Q}}\right)\right)$\\

\hline

{$\frac{5}{2}^{+}$}  &{{${^{4}P_{\frac{5}{2}}}$}} &${\frac{1}{2}}^{-}\otimes1^{+}\otimes1^{-}$ & $\mu _1+\mu _Q+\frac{2 \mu _{{q_1}}}{3}+\frac{2 \mu _{{q_2}}}{3}+\mu _{{q_3}}-\frac{\mu _{\bar{Q}}}{3}$ \\ &&${\frac{3}{2}}^{-}\otimes0^{+}\otimes1^{-}$ &$\mu _1+\mu _{{q_1}}+\mu _{{q_2}}+\mu _{\bar{Q}}$\\

&& $[{\frac{3}{2}}^{-}\otimes1^{+}]_{\frac{3}{2}}\otimes1^{-}$
&$\frac{1}{15} \left(15 \mu _1+6 \mu _Q+11 \mu _{{q_1}}+11 \mu
_{{q_2}}+6 \mu _{{q_3}}+11 \mu _{\bar{Q}}\right)$\\\cline{2-4}

 &{\multirow{1}*{${^{6}P_{\frac{5}{2}}}$}}&$[{\frac{3}{2}}^{-}\otimes1^{+}]_{\frac{5}{2}}\otimes1^{-}$  &$\frac{1}{35} \left(10 \mu _1+31 \left(\mu _Q+\mu _{{q_1}}+\mu _{{q_2}}+\mu _{{q_3}}+\mu _{\bar{Q}}\right)\right)$\\\hline

\multirow{1}{*}{$\frac{7}{2}^{+}$}  &{\multirow{1}*{${^{6}P_{\frac{7}{2}}}$}} &${\frac{3}{2}}^{-}\otimes1^{+}\otimes1^{-}$ & $\mu _1+\mu _Q+\mu _{{q_1}}+\mu _{{q_2}}+\mu _{{q_3}}+\mu _{\bar{Q}}$\\

\bottomrule[1pt]\bottomrule[1pt]

      \end{tabular}
  \end{center}
\end{table*}

\renewcommand{\arraystretch}{1.8}
\begin{table*}[htbp]
\caption{The flavor wave functions of the pentaquark states in the
diquark-triquark model with the configuration $(q_1q_2)(\bar c c
q_3)$ in the different flavor representation from ${3}_f\otimes
3_f\otimes 3_f=10_f\oplus8_{1f}\oplus8_{2f}\oplus1_f$. Here,
$\{q_1q_2\}=\frac{1}{\sqrt 2}(q_1q_2+q_2q_1)$,
$[q_1q_2]=\frac{1}{\sqrt
2}(q_1q_2-q_2q_1)$.}\label{ccbardt10-flavor}
\begin{center}
   \begin{tabular}{c|c|c} \toprule[1pt]\toprule[1pt]

$(Y, I, I_3)$ &\multicolumn{1}{c|}{ {Wave function}-$8_{1f}$} &Wave function-$8_{2f}$\\
\midrule[1pt]

$(1,\frac{1}{2},\frac{1}{2})$ &$-\sqrt{\frac{1}{3}}\{ud\}({\bar
c}cu)+\sqrt{\frac{2}{3}}\{uu\}(\bar c cd)$
& $[ud](\bar c{c}u)$ \\

$(1,\frac{1}{2},-\frac{1}{2})$  &$\sqrt{\frac{1}{3}}\{ud\}(\bar
ccd)-\sqrt{\frac{2}{3}}\{dd\}(\bar ccu)$
&  $[ud](\bar c{c}d)$\\

$(-1,\frac{1}{2},\frac{1}{2})$ &$\sqrt{\frac{1}{3}}\{us\}(\bar c
cs)-\sqrt{\frac{2}{3}}\{ss\}(\bar ccu)$
&  $[us](\bar c{c}s)$\\

$(-1,\frac{1}{2},-\frac{1}{2})$  &$\sqrt{\frac{1}{3}}\{ds\}(\bar c
cs)-\sqrt{\frac{2}{3}}\{ss\}({\bar c}cd)$
&        $[ds](\bar c cs)$\\

$ (0,1,1)$& $\sqrt{\frac{1}{3}}\{us\}(\bar c
cu)-\sqrt{\frac{2}{3}}\{uu\}(\bar ccs)$
&      $[us](\bar c {c}u)$\\

$ (0,1,0)$& $\sqrt{\frac{1}{6}}[\{us\}(\bar c cd)+\{ds\}(\bar
ccu)]-\sqrt{\frac{2}{3}}\{ud\}(\bar ccs)$
&$\frac{1}{\sqrt2}\{ [us](\bar c {c}d)+[ds](\bar c){c}u \}$ \\

$ (0,1,-1)$& $\sqrt{\frac{1}{3}}\{ds\}(\bar
ccd)-\sqrt{\frac{2}{3}}\{dd\}(\bar c cs)$
&  $[ds](\bar c {c}d)$ \\

$ (0,0,0)$&  $\sqrt{\frac{1}{2}}[ \{ds\}(\bar ccu)-(\{us\})(\bar
ccd)]$
&    $\frac{1}{\sqrt6} \{ [us]({\bar c}cd)-[ds](\bar c{c}u)-2[ud](\bar ccs) \}$\\

\midrule[1pt]

$(Y, I, I_3)$ &\multicolumn{1}{c|} {Wave function-$10_f$}
&Singlet\\
\midrule[1pt]

$(1,\frac{3}{2},\frac{3}{2})$& $\{uu\}(\bar c cu)$
&   $\frac{1}{\sqrt3} \{[us](\bar c cd)-[ds](\bar c cu)+[ud](\bar c cs) \}$\\

$(1,\frac{3}{2},\frac{1}{2})$  &$\sqrt{\frac{2}{3}}\{ud\}(\bar c cu)+\sqrt{\frac{1}{3}}\{uu\}(\bar ccd)$  \\
$(1,\frac{3}{2},-\frac{1}{2})$&  $\sqrt{\frac{2}{3}}\{ud\}(\bar c cd)+\sqrt{\frac{1}{3}}\{dd\}(\bar c cu)$  \\
$(1,\frac{3}{2},-\frac{3}{2})$&  $\{dd\}(\bar c cd)$ \\

$ (0,1,1)$& $\sqrt{\frac{2}{3}}\{us\}(\bar c cu)+\sqrt{\frac{1}{3}}\{uu\}(\bar c cs)$\\
$ (0,1,0)$& $\sqrt{\frac{1}{3}}[\{us\}(\bar c cd)+\{ds\}(\bar c cu)]+\sqrt{\frac{1}{3}}\{ud\}(\bar ccs)$ \\
$ (0,1,-1)$& $\sqrt{\frac{2}{3}}\{ds\}(\bar ccd)+\sqrt{\frac{1}{3}}\{dd\}(\bar c cs)$ \\
$ (-1,\frac{1}{2},\frac{1}{2})$& $\sqrt{\frac{1}{3}}\{ss\}(\bar c cu)+\sqrt{\frac{2}{3}}\{us\}(\bar c cs)$ \\
$ (-1,\frac{1}{2},-\frac{1}{2})$& $\sqrt{\frac{1}{3}}\{ss\}(\bar c d)+\sqrt{\frac{2}{3}}\{ds\}(\bar c cs)$ \\
$(-2,0,0)$&   $\{ss\}(\bar c cs)$   \\

\bottomrule[1pt]\bottomrule[1pt]
        \end{tabular}
  \end{center}
\end{table*}

\renewcommand{\arraystretch}{1.8}
\begin{table*}[htbp]
\caption{The magnetic moment of the pentaquark states as a
diquark-triquark state $(q_1q_2)(\bar c c q_3)$ with
$(I,I_3)=(\frac{1}{2},\frac{1}{2})$. The $8_{1f}$ and $8_{2f}$
flavor representations are from the
the${\bar3}_f\otimes3_f=1_f\oplus8_{2f}$ and ${6}_f\otimes
 3_f=10_f\oplus8_{1f}$, respectively.
%in the${3}_f\otimes 3_f=1_f\oplus8_{2f}$ flavor representation.
The third line denotes the $J^P$ of the diquark $c q_3$, the
antiquark $\bar c$, the diquark $q_1q_2$, and the orbital excitation. The
subscripts $\frac{1}{2}^-$($\frac{3}{2}^-$) represent the $J^P$ of
the triquark $\bar c c q_3$, while the subscripts $0$(or $1$) are the
total spin of the diquark $c q_3$ in the triquark $\bar c c q_3$.
The unit is the magnetic moment of the proton.}
\label{adanti8+1ccbar}
 \scriptsize
\begin{center}
   \begin{tabular}{c|c|c|c|c|c|c|c} \toprule[1pt]\toprule[1pt]

\multicolumn{8}{c}{Flavor representation-$8_{2f}$}\\
\hline &\multicolumn{2}{c|}{$^2S_{\frac{1}{2}}$
($J^P={\frac{1}{2}}^{-}$)}
&\multicolumn{1}{c|}{$^{4}S{{\frac{3}{2}}}$
(${J^P={\frac{3}{2}}^{-}}$)}
&\multicolumn{2}{c|}{${^{2}P_{\frac{1}{2}}}$
(${J^P={\frac{1}{2}}^{+}}$)}
&\multicolumn{1}{c|}{${^{4}P_{\frac{1}{2}}}$
(${J^P={\frac{1}{2}}^{+}}$)}\\\cline{2-7} \hline
 $(Y, I, I_3)$   &
${(0\otimes
\frac{1}{2})_{{\frac{1}{2}}^{-}}}\otimes0^{+}\otimes0^{+}$
      & $(1\otimes\frac{1}{2})_{{\frac{1}{2}}^{-}}\otimes0^{+}\otimes0^{+}$   & $(1\otimes \frac{1}{2})_{{\frac{3}{2}}^{-}}\otimes0^{+}\otimes0^{+}$  &  $(0\otimes \frac{1}{2})_{{\frac{1}{2}}^{-}}\otimes0^{+}\otimes1^{-}$ &$(1\otimes{\frac{1}{2}})_{{\frac{1}{2}}^{-}}\otimes 0^{+}\otimes1^{-}$
      & $(1\otimes {\frac{1}{2}}^{-})_{{\frac{3}{2}}^{-}}\otimes 0^{+}\otimes1^{-}$ \\
\hline

$(1,\frac{1}{2},\frac{1}{2})$ &-0.403&1.667&1.895&0.399&-0.291&0.921\\

\midrule[1pt]

  &\multicolumn{2}{c|}{$^{2}P_{\frac{3}{2}}$ ($J^P={\frac{3}{2}}^{+}$)} &\multicolumn{1}{c|}{$^{4}P_{\frac{3}{2}}$ ($J^P={\frac{3}{2}}^{+}$)}   &\multicolumn{1}{c|}{${^{4}P_{\frac{5}{2}}^{+}}$ (${J^P={\frac{5}{2}}^{+}}$)}   \\\cline{2-7}
  \hline
$(Y, I, I_3)$ & $(0\otimes \frac{1}{2})_{{\frac{1}{2}}^{-}}\otimes
0^{+}\otimes1^{-}$
      & $(1\otimes \frac{1}{2})_{{\frac{1}{2}}^{-}}\otimes 0^{+}\otimes1^{-}$  &  $(1\otimes \frac{1}{2})_{{\frac{3}{2}}^{-}}\otimes 0^{+}\otimes1^{-}$ & $(1\otimes \frac{1}{2})_{{\frac{3}{2}}^{-}}\otimes0^{+}\otimes1^{-}$ \\
\hline

$(1,\frac{1}{2},\frac{1}{2})$ &-0.007&2.063&1.548&2.291\\

     \bottomrule[1pt]
     %\bottomrule[1pt]
 %\end{tabular}
 % \end{center}
%\end{table*}

%\renewcommand{\arraystretch}{1.8}
%\begin{table*}[htbp]
%\caption{ The magnetic moment of the pentaquark states in the
%diquark-triquark model with the configuration $(q_1q_2)(\bar c c
%q_3)$ in the $8_{1f}$ flavor representation from ${6}_f\otimes
%3_f=10_f\oplus8_{1f}$. The third line denotes the $J^P$ of the
%triquark $\bar c c q_3$, the diquark $q_1q_2$ and orbital
%excitation. The subscripts $0$(or $1$) is the total spin of the
%diquark $c q_3$ in the triquark $\bar c c q_3$. The unit is the
%magnetic moment of the proton.}\label{ccbardt8-}
  % \scriptsize
%\begin{center}
  % \begin{tabular}{c|c|c|c|c|c|c|c} \toprule[1pt]\toprule[1pt]

\multicolumn{8}{c}{Flavor representation-$8_{1f}$}\\
\hline &\multicolumn{3}{c|}{$^{2}S_{\frac{1}{2}}$
($J^P={\frac{1}{2}}^{-}$)}
&\multicolumn{3}{c|}{${^{4}S_{\frac{3}{2}}}$
(${J^P={\frac{3}{2}}^{-}}$)}
&\multicolumn{1}{c}{$^{6}S_{\frac{5}{2}}$ ($J^P={\frac{5}{2}}^{-}$)}
\\\cline{2-8} $(Y, I, I_3)$ &
${{\frac{1}{2}}_{0}^{-}}\otimes1^{+}\otimes0^{+}$
      & ${{\frac{1}{2}}_1^{-}}\otimes1^{+}\otimes0^{+}$  &  ${{\frac{3}{2}}_1^{-}}\otimes1^{+}\otimes0^{+}$      & ${{\frac{1}{2}}_0^{-}}\otimes1^{+}\otimes0^{+}$ &${{\frac{1}{2}}_1^{-}}\otimes0^{+}\otimes0^{+}$
      & ${{\frac{3}{2}}_1^{-}}\otimes1^{+}\otimes0^{+}$  & ${{\frac{3}{2}}_1
      ^{-}}\otimes1^{+}\otimes0^{+}$
\\
\hline

$(1,\frac{1}{2},\frac{1}{2})$ &2.029&1.760&-0.947&2.439&3.246&1.137&2.842\\

     \bottomrule[1pt]

 &\multicolumn{1}{c|}{${^{2}P_{\frac{1}{2}}}$ (${J^P={\frac{1}{2}}^{+}}$)} &\multicolumn{1}{c|}{${^{4}P_{\frac{1}{2}}}$ (${J^P={\frac{1}{2}}^{+}}$)} &\multicolumn{1}{c|}{${^{2}P_{\frac{1}{2}}}$ (${J^P={\frac{1}{2}}^{+}}$)} &\multicolumn{1}{c|}{${^{4}P_{\frac{1}{2}}}$ (${J^P={\frac{1}{2}}^{+}}$)}  &\multicolumn{1}{c|}{${^{2}P_{\frac{1}{2}}}$ (${J^P={\frac{1}{2}}^{+}}$)} &\multicolumn{1}{c|}{${^{4}P_{\frac{1}{2}}}$ (${J^P={\frac{1}{2}}^{+}}$)}  \\\cline{2-8}
$(Y, I, I_3)$ & $[{{\frac{1}{2}}_0^{-}}\otimes1^{+}]_{\frac{1}{2}}\otimes1^{-}$  &  $[{{\frac{1}{2}}_0^{-}}\otimes1^{+}]_{\frac{3}{2}}\otimes1^{-}$      & $[{{\frac{1}{2}}_1^{-}}\otimes1^{+}]_{\frac{1}{2}}\otimes1^{-}$  &  $[{{\frac{1}{2}}_1^{-}}\otimes1^{+}]_{\frac{3}{2}}\otimes1^{-}$      & $[{{\frac{3}{2}}_1^{-}}\otimes1^{+}]_{\frac{1}{2}}\otimes1^{-}$  &  $[{{\frac{3}{2}}_1^{-}}\otimes1^{+}]_{\frac{3}{2}}\otimes1^{-}$ \\
\hline

$(1,\frac{1}{2},\frac{1}{2})$ &-0.130&1.082&-0.038&1.529&0.865&0.357\\

     \bottomrule[1pt]

&\multicolumn{1}{c|}{${^{2}P_{\frac{3}{2}}}$
(${J^P={\frac{3}{2}}^{+}}$)}
&\multicolumn{1}{c|}{${^{4}P_{\frac{3}{2}}}$
(${J^P={\frac{3}{2}}^{+}}$)}
&\multicolumn{1}{c|}{${^{2}P_{\frac{3}{2}}}$
(${J^P={\frac{3}{2}}^{+}}$)}
&\multicolumn{1}{c|}{${^{4}P_{\frac{3}{2}}}$
(${J^P={\frac{3}{2}}^{+}}$)}
&\multicolumn{1}{c|}{${^{2}P_{\frac{3}{2}}}$
(${J^P={\frac{3}{2}}^{+}}$)}
&\multicolumn{1}{c|}{${^{4}P_{\frac{3}{2}}}$
(${J^P={\frac{3}{2}}^{+}}$)}
&\multicolumn{1}{c}{${^{6}P_{\frac{3}{2}}}$
(${J^P={\frac{3}{2}}^{+}}$)} \\\cline{2-8}
$(Y, I, I_3)$  & $[{{\frac{1}{2}}_0^{-}}\otimes1^{+}]_{\frac{1}{2}}\otimes1^{-}$  &  $[{{\frac{1}{2}}_0^{-}}\otimes1^{+}]_{\frac{3}{2}}\otimes1^{-}$      & $[{{\frac{1}{2}}_1^{-}}\otimes1^{+}]_{\frac{1}{2}}\otimes1^{-}$  &  $[{{\frac{1}{2}}_1^{-}}\otimes1^{+}]_{\frac{3}{2}}\otimes1^{-}$      & $[{{\frac{3}{2}}_1^{-}}\otimes1^{+}]_{\frac{1}{2}}\otimes1^{-}$  &  $[{{\frac{3}{2}}_1^{-}}\otimes1^{+}]_{\frac{3}{2}}\otimes1^{-}$ &  $[{{\frac{3}{2}}_1^{-}}\otimes1^{+}]_{\frac{5}{2}}\otimes1^{-}$ \\
\hline

$(1,\frac{1}{2},\frac{1}{2})$ &2.850&2.117&2.584&2.710&-0.124&1.163&1.894\\

     \bottomrule[1pt]

   &\multicolumn{3}{c|}{${^{4}P_{\frac{5}{2}}}$ (${J^P={\frac{5}{2}}^{+}}$)} &\multicolumn{1}{c|}{${^{6}P_{\frac{5}{2}}}$ (${J^P={\frac{5}{2}}^{+}}$)}  &\multicolumn{1}{c|}{${^{6}P_{\frac{7}{2}}}$ (${J^P={\frac{7}{2}}^{+}}$)}\\\cline{2-8}

$(Y, I, I_3)$   &
$[{{\frac{1}{2}}_0^{-}}\otimes1^{+}]_{\frac{3}{2}}\otimes1^{-}$ &
$[{{\frac{1}{2}}_1^{-}}\otimes1^{+}]_{\frac{3}{2}}\otimes1^{-}$ &
$[{{\frac{3}{2}}_1^{-}}\otimes1^{+}]_{\frac{3}{2}}\otimes1^{-}$
  &  $[{{\frac{3}{2}}_1^{-}}\otimes1^{+}]_{\frac{5}{2}}\otimes1^{-}$  &  $[{{\frac{3}{2}}_1^{-}}\otimes1^{+}]_{\frac{5}{2}}\otimes1^{-}$\\
\hline
$(1,\frac{1}{2},\frac{1}{2})$ &3.259&4.069&1.960&2.753&3.666\\

  \bottomrule[1pt]
     \bottomrule[1pt]
      \end{tabular}
  \end{center}
\end{table*}

In Tables \ref{dt} and \ref{adanti8+1ccbar}, for the states with
$J^P=\frac{1}{2}^-$ and configuration $\frac{1}{2}^- \otimes 0^{+}
\otimes 0^+$ with the inner structure $(cq_3)(q_1q_2\bar{c})$ or
$(0\otimes \frac{1}{2})_{\frac{1}{2}^-}\otimes 0^{+}\otimes 0^{+}$
with the inner structure $(q_1q_2)(\bar{c}cq_3)$, the magnetic
moments are the same, which are $-0.403\mu_p$. In Tables \ref{dda8r}
and \ref{dda8l}, the magnetic moments of the
diquark-diquark-antiquark states with $J^P=\frac{1}{2}^-$ and
configuration $0^{+}\otimes 0^{+}\otimes\frac{1}{2}^-\otimes 0^+$
are also $-0.403\mu_p$. In this case, the spins of the diquark
$(cq_3)$ and diquark $(q_1q_2)$ are both $0$ in these
configurations. Moreover, there is no orbital excitation. Thus, only
the anticharm quark contributes to the magnetic moment, which leads
to the same negative magnetic moment even for the nine flavor
configurations both in the diquark-triquark model and the
diquark-diquark-antiquark model.

We collect the numerical results for the states in the $1_f\oplus
8_{2f}$ representation within the diquark-diquark-antiquark scheme
in Tables \ref{dda8r} and \ref{dda8l}, which can be compared with
the results in the diquark-triquark scheme in Tables \ref{dt}
and \ref{adanti8+1ccbar}. In these tables, the magnetic moments of
the states without P-wave excitations are the same. However, those
of the states with the P-wave excitation are slightly different. In
this case, the total spin of $q_1q_2$ is $0$. The expressions for
the magnetic moments are the same within the diquark-triquark and
the diquark-diquark-antiquark configurations as illustrated in
Tables \ref{analysis8+1} and \ref{analysis3}.

The expressions of the magmatic moments of some states in the
$8_{2f}\oplus 10_f$ representations are also the same in different
models. For example, the magnetic moments of the states with
$J^P=\frac{1}{2}^{-}$ and configuration $\frac{1}{2}^- \otimes
0^{+}\otimes {0}^{+}$ in the diquark-triquark model are the same as
those of the states with $J^P=\frac{1}{2}^{-}$ and configuration
$0^{+}\otimes 1^{+}\otimes{\frac{1}{2}^{-}}\otimes{0^{+}}$ in the
diquark-diquark-antiquark model. For the states with P-wave
excitation, their different magnetic moments are due to the
different  P-wave excitation modes and different masses of the
related clusters.

For the molecular-type pentaquark states, their expressions can also
be obtained through the interexchange of $\mu_Q$ and $\mu_{\bar Q}$
in Tables \ref{analysis8+1}. Therefore, the numerical results are
usually different. However, for some states without orbital
excitation, the analytic expressions are symmetric under
interexchange of $\mu_Q$ and $\mu_{\bar Q}$ for some special
configurations. Then, the magnetic moments of the molecular-type
pentaquark states are the same as those of the diquark-triquark-type
states {with the configuration $(c q_3)(\bar c q_1q_2)$. For
example, the results of the molecular-type states with
$J^P=\frac{5}{2}^{-}$ and configuration $\frac{3}{2}^{+} \otimes
1^{-}\otimes {0^+}$ are the same as those of the diquark-triquark-type states with the configuration $\frac{3}{2}^{-} \otimes
1^{+}\otimes {0^+}$. Moreover, the magnetic moments of the states
with $J^P=\frac{5}{2}^{-}$ are the same in both the
diquark-diquark-antiquark model and diqaurk-triquark model as
illustrated in Sec .\ref{sec4}. Thus, their magnetic moments are the
same in the above three models.
%The same conclusions hold for the states with $J^P=\frac{5}{2}^-$ are also the same.
%$(\frac{1}{2}^{-} \otimes1^{+})_{\frac{3}{2}^-}\otimes 1^{+}\otimes{0^+}$

\section{Summary}
\label{sec6}
%%%%%%%%%%%%%%%%%%%%

Searching for exotic states is an intriguing issue in hadron
physics. In 2015, two candidates of the hidden-charm pentaquark were
reported by LHCb \cite{Aaij:2015tga}, which were confirmed very
recently by LHCb by performing a model-independent analysis
\cite{Aaij:2016phn}. The observed $P_c(4380)$ and $P_c(4450)$ states
have stimulated extensive discussions of their inner structures. As
summarized in Sec. \ref{sec1}, there exist several speculations of
the inner structures of $P_c(4380)$ and $P_c(4450)$, which include
the molecular state scheme, diquark-diquark-antiquark scheme and the
diquark-triquark scheme. How to distinguish these possible
configurations of the $P_c$ states becomes a crucial task.

In this work, we propose that the magnetic moments of the
hidden-charm pentaquarks can be employed to achieve this aim. We
have calculated the magnetic moments of the hidden-charm pentaquark
states with $J^P=\frac{1}{2}^{\pm}, \frac{3}{2}^{\pm},
\frac{5}{2}^{\pm}$ and $\frac{7}{2}^{+}$, where we consider
different configurations of the hidden-charm pentaquark states.

In fact, the magnetic moments of the states with the same quantum
numbers are usually different in the above three models. We take the
state with $J^P=\frac{7}{2}^+$ and the isospin $(I,
I_z)=(\frac{1}{2},\frac{1}{2})$ in the $8_{1f}$ representation as an
example. Its magnetic moment is $2.945\mu_p$, $3.651\mu_p$,
$4.496\mu_p$, $3.068\mu_p$ and $3.666\mu_p$ when it is a molecular-type state, a diquark-diquark-antiquark-type state with the $\rho$
excitation mode, a diquark-diquark-antiquark-type state with the
$\lambda$ excitation mode, a diquark-triquark-type state with the
configurations $(c q_3)(\bar c q_1 q_2)$ and $(q_1 q_2)(\bar c  c
q_3)$, respectively.

Even within the same model, one set of the $J^P$ quantum numbers may
correspond to several states with different configurations. Their
magnetic moments are also different. Let's take the magnetic moments
of the hidden charm molecular states composed of $\Sigma_c \bar
D^{\ast}$ in the $8_{1f}$ representation with $J^P=\frac{3}{2}^{+}$
and the isospin $(I, I_z)=(\frac{1}{2},\frac{1}{2})$ in Table
\ref{pcm} as an example. The magnetic moments of the pentaquark
states with different configurations
$[\frac{1}{2}^{+}\otimes1^{-}]_{\frac{1}{2}}\otimes 1^{-}$ and
$[\frac{1}{2}^{+}\otimes1^{-}]_{\frac{3}{2}}\otimes 1^{-}$ are very
different, which are $-0.740\mu_p$ and $1.041\mu_p$, respectively.

In the diquark-diquark-antiquark model, we choose the states with
the isospin $(I, I_z)=(\frac{1}{2},\frac{1}{2})$ in the $8_{1f}$
representation as an example. In Tables \ref{dda8r} and \ref{dda8l},
the magnetic moments of the $J^P=\frac{3}{2}^+$ states with the
p-wave $\rho$ excitation mode and configurations
$(({1^+}\otimes{1^+})_0\otimes {\frac{1}{2}^-})_{\frac{1}{2}}\otimes
1^-$ and $(({1^+}\otimes{1^+})_1\otimes
{\frac{1}{2}^{-}})_{\frac{1}{2}}\otimes 1^-$ are $0.405{\mu}_p$ and
$2.025\mu_p$, respectively, while the corresponding numerical results
of these two configurations with the $\lambda$ excitation mode are
$1.250\mu_p$ and $2.870\mu_p$, respectively. The results of the
states with the same p-wave excitation mode are different if their
configurations are different. Moreover, the different p-wave
excitation modes also lead to the different numerical results.
Actually, the magnetic moments of the $\frac{3}{2}^{+}$ and
$\frac{5}{2}^{+}$ $P_c$ states in the $8_{1f}$ representation with
the isospin $(I, I_z)=(\frac{1}{2},\frac{1}{2})$ and the $\lambda$
excitation mode are larger than those of the states with the $\rho$
excitation mode except the configuration ${((1^+\otimes1^+)_2\otimes
\frac{1}{2}^- })_{\frac{5}{2}}\otimes1^-$ as illustrated in Sec.
\ref{sec4}.

This also happens in the diquark-triquark model. For example, in
Table \ref{dt}, the magnetic moments of the $\frac{1}{2}^+$ states
as a $(c q_3)(q_1q_2\bar c)$ state in the $8_{2f}$ flavor
representation with configurations $(\frac{1}{2}^{-}\otimes
1^+)_{\frac{1}{2}}\otimes 1^-$ and $(\frac{1}{2}^{-}\otimes
1^+)_{\frac{3}{2}}\otimes 1^-$ are $-0.384\mu_p$ and $0.967\mu_p$
respectively.

Our results clearly show that the magnetic moments of the
hidden-charm pentaquarks with different configurations are very
different. In other words, the experimental measurement of the
magnetic moment of the hidden-charm pentaquark indeed can help
distinguish their inner structure.

The $P_c$ states can be produced through either the photo-production
or electro-production process $\gamma^{(*)} P\to P_c \to J/\psi P$.
In order to extract the magnetic moments of the $P_c$ states, one
could study one more challenging process $\gamma^{(*)} P\to P_c \to
P_c \gamma\to J/\psi P \gamma$. A very similar process $\gamma^{(*)}
P\to \Delta^+ \to \Delta^+ \gamma\to \pi^0 P \gamma$ was used to
extract the magnetic moment of the $\Delta$ resonance
\cite{Pascalutsa:2004je,Pascalutsa:2007wb}. The total and differential cross sections may
be sensitive to the magnetic moment of the hidden-charm pentaquark,
which might be extracted by comparing theoretical predictions with
the measured cross sections. We are looking forward to future
experimental progress along these directions.

%%%%%%%%%%%%%%%%%%%%%%%%%%%%%
\section*{Acknowledgments}
%%%%%%%%%%%%%%%%%%%%%%%%%%%%%%

This project is supported by National Natural Science Foundation of
China under Grants No. 11222547, No. 11175073, No. 11575008, and 973
program. X.L. is also supported by National Program for Support of Top-notch Youth Professionals.


\begin{thebibliography}{99}

%\cite{Aaij:2015tga}
\bibitem{Aaij:2015tga}
  R.~Aaij {\it et al.} (LHCb Collaboration),
  Observation of $J/\psi p$ Resonances Consistent with Pentaquark States in $\Lambda_b^0\to J/\psi K^-p$ Decays,
  Phys.\ Rev.\ Lett.\  {\bf 115}, 072001 (2015).
  %%CITATION = ARXIV:1507.03414;%%
  %68 citations counted in INSPIRE as of 18 Oct 2015
%\cite{Aaij:2015tga}


%\cite{Aaij:2016phn}
\bibitem{Aaij:2016phn} 
  R.~Aaij {\it et al.} (LHCb Collaboration),
  Model-independent evidence for $J/\psi p$ contributions to $\Lambda_b^0\to J/\psi p K^-$ decays,
  Phys.\ Rev.\ Lett.\  {\bf 117}, 082002 (2016).

%  doi:10.1103/PhysRevLett.117.082002
 % [arXiv:1604.05708 [hep-ex]].
  %%CITATION = doi:10.1103/PhysRevLett.117.082002;%%
  %20 citations counted in INSPIRE as of 21 Nov 2016  %%CITATION = ARXIV:1604.05708;%%


%\cite{Yang:2011wz}
\bibitem{Yang:2011wz}
  Z.~C.~Yang, Z.~F.~Sun, J.~He, X.~Liu and S.~L.~Zhu,
  The possible hidden-charm molecular baryons composed of anti-charmed meson and charmed baryon,
  Chin.\ Phys.\ C {\bf 36}, 6 (2012).
  %%CITATION = ARXIV:1105.2901;%%
  %8 citations counted in INSPIRE as of 17 juin 2015

%\cite{Wu:2010jy}
\bibitem{Wu:2010jy}
  J.~J.~Wu, R.~Molina, E.~Oset and B.~S.~Zou,
  Prediction of narrow $N^*$ and $\Lambda^*$ resonances with hidden charm above 4 GeV,
  Phys.\ Rev.\ Lett.\  {\bf 105}, 232001 (2010).
  %doi:10.1103/PhysRevLett.105.232001
  %%CITATION = doi:10.1103/PhysRevLett.105.232001;%%
  %86 citations counted in INSPIRE as of 11 Jan 2016

%\cite{Wu:2010vk}
\bibitem{Wu:2010vk}
  J.~J.~Wu, R.~Molina, E.~Oset and B.~S.~Zou,
  Dynamically generated $N^{*}$ and $\Lambda^*$ resonances in the hidden charm sector around 4.3 GeV,
  Phys.\ Rev.\ C {\bf 84}, 015202 (2011).
  %doi:10.1103/PhysRevC.84.015202
  
  %%CITATION = doi:10.1103/PhysRevC.84.015202;%%
  %64 citations counted in INSPIRE as of 11 Jan 2016

%\cite{Wang:2011rga}
\bibitem{Wang:2011rga}
  W.~L.~Wang, F.~Huang, Z.~Y.~Zhang and B.~S.~Zou,
  $\Sigma_c \bar{D}$ and $\Lambda_c \bar{D}$ states in a chiral quark model,
  Phys.\ Rev.\ C {\bf 84}, 015203 (2011).
  %doi:10.1103/PhysRevC.84.015203

  %%CITATION = doi:10.1103/PhysRevC.84.015203;%%
  %27 citations counted in INSPIRE as of 11 Jan 2016


%\cite{Chen:2016qju}
\bibitem{Chen:2016qju} 
  H.~X.~Chen, W.~Chen, X.~Liu and S.~L.~Zhu,
  The hidden-charm pentaquark and tetraquark states,
  Phys.\ Rept.\  {\bf 639}, 1 (2016).
%  doi:10.1016/j.physrep.2016.05.004
  %[arXiv:1601.02092 [hep-ph]].
  %%CITATION = doi:10.1016/j.physrep.2016.05.004;%%
  %96 citations counted in INSPIRE as of 21 Nov 2016
%\cite{Karliner:2015ina}
\bibitem{Karliner:2015ina}
  M.~Karliner and J.~L.~Rosner,
  New Exotic Meson and Baryon Resonances from Doubly-Heavy Hadronic Molecules,
  Phys.\ Rev.\ Lett.\  {\bf 115}, 122001 (2015)
 % doi:10.1103/PhysRevLett.115.122001

  %%CITATION = doi:10.1103/PhysRevLett.115.122001;%%
  %42 citations counted in INSPIRE as of 22 Apr 2016

%\cite{Chen:2015loa}
\bibitem{Chen:2015loa}
  R.~Chen, X.~Liu, X.~Q.~Li and S.~L.~Zhu,
  Identifying exotic hidden-charm pentaquarks,
  Phys.\ Rev.\ Lett.\  {\bf 115},132002 (2015).
 % doi:10.1103/PhysRevLett.115.132002
  %[arXiv:1507.03704 [hep-ph]].
  %%CITATION = doi:10.1103/PhysRevLett.115.132002;%%
  %50 citations counted in INSPIRE as of 20 Apr 2016


%\cite{Chen:2015moa}
\bibitem{Chen:2015moa}
  H.~X.~Chen, W.~Chen, X.~Liu, T.~G.~Steele and S.~L.~Zhu,
  Towards exotic hidden-charm pentaquarks in QCD,
  Phys.\ Rev.\ Lett.\  {\bf 115}, 172001 (2015).
  %doi:10.1103/PhysRevLett.115.172001
%  [arXiv:1507.03717 [hep-ph]].
  %%CITATION = doi:10.1103/PhysRevLett.115.172001;%%
  %43 citations counted in INSPIRE as of 20 Apr 2016


  %\cite{Huang:2015uda}
\bibitem{Huang:2015uda}
  H.~Huang, C.~Deng, J.~Ping and F.~Wang,
  Possible pentaquarks with heavy quarks,
  arXiv:1510.04648 [hep-ph].
  %%CITATION = ARXIV:1510.04648;%%
  %6 citations counted in INSPIRE as of 07 Jan 2016

%\cite{Yang:2015bmv}
\bibitem{Yang:2015bmv}
  G.~Yang and J.~Ping,
  The Structure of Pentaquarks $P_c^+$ in the Chiral Quark Model,
  arXiv:1511.09053 [hep-ph].
  %%CITATION = ARXIV:1511.09053;%%
  %6 citations counted in INSPIRE as of 29 Apr 2016

  %\cite{Roca:2015dva}
\bibitem{Roca:2015dva}
  L.~Roca, J.~Nieves and E.~Oset,
  LHCb pentaquark as a $\bar{D}^*\Sigma_c-\bar{D}^*\Sigma_c^*$ molecular state,
  Phys.\ Rev.\ D {\bf 92}, 094003 (2015)
%  doi:10.1103/PhysRevD.92.094003
 % [arXiv:1507.04249 [hep-ph]].
  %%CITATION = doi:10.1103/PhysRevD.92.094003;%%
  %49 citations counted in INSPIRE as of 20 Apr 2016


%\cite{He:2015cea}
\bibitem{He:2015cea}
  J.~He,
  $\bar{D}\Sigma^*_c$ and $\bar{D}^*\Sigma_c$ interactions and the LHCb hidden-charmed pentaquarks,
  Phys.\ Lett.\ B {\bf 753}, 547 (2016).
%  doi:10.1016/j.physletb.2015.12.071
 % [arXiv:1507.05200 [hep-ph]].
  %%CITATION = doi:10.1016/j.physletb.2015.12.071;%%
  %43 citations counted in INSPIRE as of 20 Apr 2016


%\cite{Burns:2015dwa}
\bibitem{Burns:2015dwa}
  T.~J.~Burns,
  Phenomenology of P$_{c}$(4380)$^{+}$, P$_{c}$(4450)$^{+}$ and related states,
  Eur.\ Phys.\ J.\ A {\bf 51},152 (2015).
%  doi:10.1140/epja/i2015-15152-6
  %[arXiv:1509.02460 [hep-ph]].
  %%CITATION = doi:10.1140/epja/i2015-15152-6;%%
  %14 citations counted in INSPIRE as of 07 Jan 2016

%\cite{Lu:2016nnt}
\bibitem{Lu:2016nnt} 
  Q.~F.~LŸ and Y.~B.~Dong,
 Strong decay mode $J/\psi p$ of hidden charm pentaquark states $P_c^+(4380)$ and $P_c^+(4450)$ in $\Sigma_c \bar{D}^*$ molecular scenario,
 Phys.\ Rev.\ D {\bf 93}, 074020 (2016).
 % doi:10.1103/PhysRevD.93.074020
  %[arXiv:1603.00559 [hep-ph]].
  %%CITATION = doi:10.1103/PhysRevD.93.074020;%%
  %11 citations counted in INSPIRE as of 21 Nov 2016  %2 citations counted in INSPIRE as of 20 Apr 2016

 %\cite{Chen:2016heh}
\bibitem{Chen:2016heh} 
  R.~Chen, X.~Liu and S.~L.~Zhu,
Hidden-charm molecular pentaquarks and their charm-strange partners,
  Nucl.\ Phys.\ A {\bf 954}, 406 (2016).
%  doi:10.1016/j.nuclphysa.2016.04.012
  %[arXiv:1601.03233 [hep-ph]].
  %%CITATION = doi:10.1016/j.nuclphysa.2016.04.012;%%
  %14 citations counted in INSPIRE as of 21 Nov 2016
  
  
  
  %\cite{Tazimi:2016hsv}
\bibitem{Tazimi:2016hsv} 
  M.~Monemzadeh, N.~Tazimiand and S.~Babaghodrat,
Calculating Masses of Pentaquarks Composed of Baryons and Mesons,
  Adv.\ High Energy Phys.\  {\bf 2016}, 6480926 (2016).
 % doi:10.1155/2016/6480926
  %[arXiv:1601.00642 [hep-ph]].
  %%CITATION = doi:10.1155/2016/6480926;%%
  %1 citations counted in INSPIRE as of 21 Nov 2016

%\cite{Feijoo:2015kts}
\bibitem{Feijoo:2015kts} 
  A.~Feijoo, V.~K.~Magas, A.~Ramos and E.~Oset,
A hidden-charm $S=-1$ pentaquark from the decay of $\varLambda _b$ into $J/\psi \eta \varLambda$ states,
  Eur.\ Phys.\ J.\ C {\bf 76}, 446 (2016).
%  doi:10.1140/epjc/s10052-016-4302-7
  %[arXiv:1512.08152 [hep-ph]].
  %%CITATION = doi:10.1140/epjc/s10052-016-4302-7;%%
  %10 citations counted in INSPIRE as of 21 Nov 2016





  %\cite{Kahana:2015tkb}
\bibitem{Kahana:2015tkb}
  D.~E.~Kahana and S.~H.~Kahana,
  LHCb $P_c^+$ Resonances as Molecular States,
  arXiv:1512.01902 [hep-ph].
  %%CITATION = ARXIV:1512.01902;%%
  %3 citations counted in INSPIRE as of 20 Apr 2016

 %\cite{Shen:2016tzq}
\bibitem{Shen:2016tzq} 
  C.~W.~Shen, F.~K.~Guo, J.~J.~Xie and B.~S.~Zou,
Disentangling the hadronic molecule nature of the $P_c(4380)$ pentaquark-like structure,
  Nucl.\ Phys.\ A {\bf 954}, 393 (2016).
%  doi:10.1016/j.nuclphysa.2016.04.034
  %[arXiv:1603.04672 [hep-ph]].
  %%CITATION = doi:10.1016/j.nuclphysa.2016.04.034;%%
  %6 citations counted in INSPIRE as of 21 Nov 2016 

%\cite{Wang:2015qlf}
\bibitem{Wang:2015qlf} 
  G.~J.~Wang, L.~Ma, X.~Liu and S.~L.~Zhu,
 Strong decay patterns of the hidden-charm pentaquark states $P_c(4380)$ and $P_c(4450)$,
  Phys.\ Rev.\ D {\bf 93}, 034031 (2016)
%  doi:10.1103/PhysRevD.93.034031
  %[arXiv:1511.04845 [hep-ph]].
  %%CITATION = doi:10.1103/PhysRevD.93.034031;%%
  %11 citations counted in INSPIRE as of 21 Nov 2016


  %\cite{Maiani:2015vwa}
\bibitem{Maiani:2015vwa}
  L.~Maiani, A.~D.~Polosa and V.~Riquer,
  The New Pentaquarks in the Diquark Model,
  Phys.\ Lett.\ B {\bf 749}, 289 (2015).
%  doi:10.1016/j.physletb.2015.08.008
 % [arXiv:1507.04980 [hep-ph]].
  %%CITATION = doi:10.1016/j.physletb.2015.08.008;%%
  %33 citations counted in INSPIRE as of 07 Jan 2016




  %\cite{Anisovich:2015cia}
\bibitem{Anisovich:2015cia}
  V.~V.~Anisovich, M.~A.~Matveev, J.~Nyiri, A.~V.~Sarantsev and A.~N.~Semenova,
  Pentaquarks and resonances in the $pJ/\psi$ spectrum,
  arXiv:1507.07652 [hep-ph].
  %%CITATION = ARXIV:1507.07652;%%
  %23 citations counted in INSPIRE as of 07 Jan 2016



%\cite{Wang:2015epa}
\bibitem{Wang:2015epa} 
  Z.~G.~Wang,
Analysis of $P_c(4380)$ and $P_c(4450)$ as pentaquark states in the diquark model with QCD sum rules,
  Eur.\ Phys.\ J.\ C {\bf 76}, 70 (2016).
  %doi:10.1140/epjc/s10052-016-3920-4
  %[arXiv:1508.01468 [hep-ph]].
  %%CITATION = doi:10.1140/epjc/s10052-016-3920-4;%%
  %40 citations counted in INSPIRE as of 21 Nov 2016

 %\cite{Li:2015gta}
\bibitem{Li:2015gta}
  G.~N.~Li, X.~G.~He and M.~He,
Some Predictions of Diquark Model for Hidden Charm Pentaquark Discovered at the LHCb,
  JHEP {\bf 1512} (2015) 128.
%  doi:10.1007/JHEP12(2015)128
  %[arXiv:1507.08252 [hep-ph]].
  %%CITATION = doi:10.1007/JHEP12(2015)128;%%
  %40 citations counted in INSPIRE as of 21 Nov 2016

  %\cite{Lebed:2015tna}
\bibitem{Lebed:2015tna}
  R.~F.~Lebed,
  The Pentaquark Candidates in the Dynamical Diquark Picture,
  Phys.\ Lett.\ B {\bf 749}, 454 (2015).
%  doi:10.1016/j.physletb.2015.08.032
 % [arXiv:1507.05867 [hep-ph]].
  %%CITATION = doi:10.1016/j.physletb.2015.08.032;%%
  %34 citations counted in INSPIRE as of 07 Jan 2016

  %\cite{Zhu:2015bba}
\bibitem{Zhu:2015bba} 
  R.~Zhu and C.~F.~Qiao,
  %``Pentaquark states in a diquarkÐtriquark model,''
  Phys.\ Lett.\ B {\bf 756}, 259 (2016).
  %doi:10.1016/j.physletb.2016.03.022
  %[arXiv:1510.08693 [hep-ph]].
  %%CITATION = doi:10.1016/j.physletb.2016.03.022;%%
  %26 citations counted in INSPIRE as of 21 Nov 2016
  
  
  
  %\cite{Chen:2015sxa}
\bibitem{Chen:2015sxa} 
  H.~X.~Chen, L.~S.~Geng, W.~H.~Liang, E.~Oset, E.~Wang and J.~J.~Xie,
Looking for a hidden-charm pentaquark state with strangeness $S=-1$ from $\Xi_b^-$ decay into $ J/\Psi K^-\Lambda$,
  Phys.\ Rev.\ C {\bf 93}, 065203 (2016).
 % doi:10.1103/PhysRevC.93.065203
  %[arXiv:1510.01803 [hep-ph]].
  %%CITATION = doi:10.1103/PhysRevC.93.065203;%%
  %16 citations counted in INSPIRE as of 21 Nov 2016 

  %\cite{Lebed:2015dca}
\bibitem{Lebed:2015dca}
  R.~F.~Lebed,
  Do the $P_c^+$ Pentaquarks Have Strange Siblings?
  Phys.\ Rev.\ D {\bf 92}, 114030 (2015).
 % doi:10.1103/PhysRevD.92.114030
 % [arXiv:1510.06648 [hep-ph]].
  %%CITATION = doi:10.1103/PhysRevD.92.114030;%%
  %5 citations counted in INSPIRE as of 07 Jan 2016

%\cite{Lu:2015fva}
\bibitem{Lu:2015fva} 
  Q.~F.~LŸ, X.~Y.~Wang, J.~J.~Xie, X.~R.~Chen and Y.~B.~Dong,
 Neutral hidden charm pentaquark states $P_c^0(4380)$ and $P_c^0(4450)$ in $\pi^-p \to J/\psi n$ reaction,
  Phys.\ Rev.\ D {\bf 93}, 034009 (2016).
  %doi:10.1103/PhysRevD.93.034009
  %[arXiv:1510.06271 [hep-ph]].
  %%CITATION = doi:10.1103/PhysRevD.93.034009;%%
  %17 citations counted in INSPIRE as of 21 Nov 2016

\bibitem{Liu:2015fea}
  X.~H.~Liu, Q.~Wang and Q.~Zhao,
  Understanding the newly observed heavy pentaquark candidates,
  Phys.\ Lett.\ B {\bf 757}, 231 (2016).
%  doi:10.1016/j.physletb.2016.03.089
  %[arXiv:1507.05359 [hep-ph]].
  %%CITATION = doi:10.1016/j.physletb.2016.03.089;%%
  %43 citations counted in INSPIRE as of 20 Apr 2016

  %\cite{Mikhasenko:2015vca}
\bibitem{Mikhasenko:2015vca}
  M.~Mikhasenko,
  A triangle singularity and the LHCb pentaquarks,
  arXiv:1507.06552 [hep-ph].
  %%CITATION = ARXIV:1507.06552;%%
  %35 citations counted in INSPIRE as of 20 Apr 2016

    %\cite{Guo:2015umn}
\bibitem{Guo:2015umn}
  F.~K.~Guo, U.~G.~Mei§ner, W.~Wang and Z.~Yang,
  How to reveal the exotic nature of the P$_c$(4450),
  Phys.\ Rev.\ D {\bf 92}, 071502 (2015).
  %doi:10.1103/PhysRevD.92.071502
%  [arXiv:1507.04950 [hep-ph]].
  %%CITATION = doi:10.1103/PhysRevD.92.071502;%%
  %47 citations counted in INSPIRE as of 20 Apr 2016


 %\cite{Scoccola:2015nia}
\bibitem{Scoccola:2015nia}
  N.~N.~Scoccola, D.~O.~Riska and M.~Rho,
  Pentaquark candidates $P_c^+$(4380) and $P_c^+$(4450) within the soliton picture of baryons,
  Phys.\ Rev.\ D {\bf 92}, 051501 (2015).
  %doi:10.1103/PhysRevD.92.051501
  %[arXiv:1508.01172 [hep-ph]].
  %%CITATION = doi:10.1103/PhysRevD.92.051501;%%
  %20 citations counted in INSPIRE as of 20 janv. 2016

%\cite{Liu:2003ab}
\bibitem{Liu:2003ab}
  Y.~R.~Liu, P.~Z.~Huang, W.~Z.~Deng, X.~L.~Chen and S.~L.~Zhu,
  Pentaquark magnetic moments in different models,
  Phys.\ Rev.\ C {\bf 69}, 035205 (2004).
%  doi:10.1103/PhysRevC.69.035205
%  [hep-ph/0312074].
  %%CITATION = doi:10.1103/PhysRevC.69.035205;%%
  %43 citations counted in INSPIRE as of 07 Jan 2016

  %\cite{Agashe:2014kda}
\bibitem{Agashe:2014kda}
  K.~A.~Olive {\it et al.} [Particle Data Group Collaboration],
  Review of Particle Physics,
  Chin.\ Phys.\ C {\bf 38}, 090001 (2014).
%  doi:10.1088/1674-1137/38/9/090001
  %%CITATION = doi:10.1088/1674-1137/38/9/090001;%%
  %2687 citations counted in INSPIRE as of 07 Jan 2016


  %\cite{Ebert:2010af}
\bibitem{Ebert:2010af}
  D.~Ebert, R.~N.~Faustov and V.~O.~Galkin,
  Masses of tetraquarks with open charm and bottom,
  Phys.\ Lett.\ B {\bf 696}, 241 (2011).
%  doi:10.1016/j.physletb.2010.12.033
 % [arXiv:1011.2677 [hep-ph]].
  %%CITATION = doi:10.1016/j.physletb.2010.12.033;%%
  %9 citations counted in INSPIRE as of 07 Jan 2016


%\cite{Pascalutsa:2004je}
\bibitem{Pascalutsa:2004je} 
  V.~Pascalutsa and M.~Vanderhaeghen,
Magnetic moment of the Delta(1232)-resonance in chiral effective field theory,
  Phys.\ Rev.\ Lett.\  {\bf 94}, 102003 (2005).
%  doi:10.1103/PhysRevLett.94.102003
  %[nucl-th/0412113].
  %%CITATION = doi:10.1103/PhysRevLett.94.102003;%%
  %51 citations counted in INSPIRE as of 21 Nov 2016 
%\cite{Pascalutsa:2007wb}
\bibitem{Pascalutsa:2007wb} 
  V.~Pascalutsa and M.~Vanderhaeghen,
Chiral effective-field theory in the Delta(1232) region. II. Radiative pion photoproduction,
  Phys.\ Rev.\ D {\bf 77}, 014027 (2008).
 % doi:10.1103/PhysRevD.77.014027
  %[arXiv:0709.4583 [hep-ph]].
  %%CITATION = doi:10.1103/PhysRevD.77.014027;%%
  %19 citations counted in INSPIRE as of 21 Nov 2016



\end{thebibliography}
\end{document}